\documentclass[a4paper,11pt]{article}

\pdfoutput=1

\usepackage[utf8]{inputenc}
\usepackage{amsmath}
\usepackage{amssymb}
\usepackage{amsfonts}
\usepackage{mathrsfs}
\usepackage{graphicx}
\usepackage{booktabs,adjustbox}
\usepackage{caption}
\usepackage{slashed}
\usepackage{verbatim}
\usepackage{float}
\usepackage{setspace}
\usepackage{subfig}
\usepackage{jheppub}

\usepackage{color}
\usepackage{ulem}

\usepackage{bookmark}
\usepackage{wasysym}
\usepackage{multirow}
\usepackage[yyyymmdd]{datetime}

\makeatletter
\newcommand{\thickhline}{%
	\noalign {\ifnum 0=`}\fi \hrule height 1pt
	\futurelet \reserved@a \@xhline
}
\makeatother

\usepackage[utf8]{inputenc}
\usepackage{float}

\allowdisplaybreaks

\title{\boldmath   Neutrino Magnetic Moments Meet Precision $N_{\rm eff}$ Measurements}

\author[a]{Shao-Ping Li}
\author[a]{and Xun-Jie Xu}

\affiliation[a]{Institute of High Energy Physics, Chinese Academy of Sciences, Beijing 100049, China}

\emailAdd{spli@ihep.ac.cn}
\emailAdd{xuxj@ihep.ac.cn}  

\abstract{
In the early universe, Dirac neutrino magnetic moments due to their chirality-flipping nature could lead to thermal production of right-handed neutrinos, which would make a significant contribution to the effective neutrino number, $N_{\rm eff}$. 
We present in this paper a dedicated computation of the  neutrino chirality-flipping rate 
in the thermal plasma.
With a careful and consistent treatment of soft scattering and the plasmon effect in finite temperature field theories, 
we find that neutrino magnetic moments above $2.7\times 10^{-12}\mu_B$ have been excluded by current CMB and BBN measurements of $N_{\rm eff}$, 
assuming flavor-universal and diagonal magnetic moments for all three generation of neutrinos.  
This limit is stronger than the latest bounds from  XENONnT and LUX-ZEPLIN experiments and comparable with those from stellar cooling considerations.
}

\begin{document}
	\maketitle
	\flushbottom

\section{Introduction}
\label{sec:intro}
Neutrino magnetic moments (NMM) serve as  a 
window to probe new physics beyond the standard model (SM)~\cite{Giunti:2014ixa}.
For Dirac neutrinos, the SM contribution to NMM is suppressed by neutrino masses and hence extremely small, $\mu_\nu\sim 10^{-20}\mu_B$\footnote{The Bohr magneton $\mu_B$ is defined as $\mu_B=e/2m_e\approx 0.296\ {\rm MeV}^{-1}$.}~\cite{Fujikawa:1980yx,Pal:1981rm, Shrock:1982sc}.
For Majorana neutrinos, they are further suppressed by a cancellation similar to the Glashow-Iliopoulos-Maiani (GIM) mechanism~\cite{Glashow:1970gm}. 
In new physics models, the magnitude of NMM can be potentially enhanced and experimentally  accessible~\cite{Voloshin:1987qy,Barr:1990um,Barr:1990dm,Babu:1990hu,Babu:1992vq,Lindner:2017uvt,Xu:2019dxe,Babu:2020ivd}.



Large NMM have motivated extensive searches in laboratories as well as astrophysical and cosmological observations. Laboratory searches are mostly based on the measurement of low-energy neutrino scattering~\cite{Grimus:2002vb,Canas:2015yoa,Borexino:2017fbd,Huang:2018nxj,Brdar:2020quo,Coloma:2022umy,Schwemberger:2022fjl,Ye:2021zso,Yue:2021vjg,Miranda:2021kre,Akhmedov:2022txm, AtzoriCorona:2022jeb,Khan:2022bel, Li:2022bqr,A:2022acy}. Astrophysical bounds from stellar energy loss, though vulnerable to uncertainties of astrophysical models, have long been stronger than laboratory ones~\cite{Raffelt1996}. It is noteworthy, however, that the latest laboratory bounds, $\mu_\nu<6.3 \times 10^{-12} \mu_B$ from XENONnT~\cite{XENON:2022mpc} and $\mu_\nu<6.2 \times 10^{-12} \mu_B$ from LUX-ZEPLIN (LZ)~\cite{LZ:2022ufs}, are approaching the astrophysical level,  $\mu_\nu< 2.2 \times 10^{-12}\mu_B$ from Ref.~\cite{Diaz:2019kim} or $\mu_\nu< 1.5 \times 10^{-12}\mu_B$ from Ref.~\cite{Capozzi:2020cbu}.




Cosmological observables such as the effective neutrino number, $N_{\rm eff}$, can also be used to constrain NMM. For Dirac neutrinos, NMM could flip the chirality and thermalize right-handed neutrinos, which would contribute to  $N_{\rm eff}$ significantly. 
Under this line of thought, previous studies derived cosmological bounds on NMM, $\mu_\nu<\mathcal{O}(1) \times 10^{-11}\mu_B$~\cite{Morgan:1981zy,Elmfors:1997tt} and $\mu_\nu<2.9 \times 10^{-10}\mu_B$~\cite{Ayala:1999xn}. Alternatively, one may consider Majorana neutrinos which can only possess transition magnetic moments. In this case, no additional species are produced but large NMM could modify  the SM neutrino decoupling at the MeV epoch.  However, the cosmological constraint on this scenario is found to be weak~\cite{Vassh:2015yza}. 

Future CMB experiments such as CMB-S4~\cite{CMB-S4:2016ple}, SPT-3G~\cite{SPT-3G:2014dbx}, and Simons Observatory~\cite{SimonsObservatory:2019qwx} will be able to improve the measurement of $N_{\rm eff}$ substantially. This will allow for a robust test of various new physics that could potentially modify $N_{\rm eff}$~\cite{Boehm:2012gr,Kamada:2015era,deSalas:2016ztq,Kamada:2018zxi,Huang:2017egl,Escudero:2018mvt,Borah:2018gjk,Depta:2019lbe,Escudero:2020dfa,Abazajian:2019oqj,Luo:2020sho,Borah:2020boy,Adshead:2020ekg, Luo:2020fdt,Hufnagel:2021pso,Li:2021okx}.  
In this work, we present a timely investigation into the cosmological constraints on NMM and find that future $N_{\rm eff}$ measurements will be able to probe $\mu_{\nu}$ down to a level lower than the  XENONnT and LZ bounds and comparable to the astrophysical one.


Our work contains a careful calculation of the chirality-flipping rate, which was not treated consistently when an infrared (IR) divergence is involved in previous studies~\cite{Morgan:1981zy,Elmfors:1997tt,Ayala:1999xn}.  
In  Ref.~\cite{Morgan:1981zy},  the  chirality-flipping rate was obtained from a straightforward computation with a naive cut for the momentum transfer and the result exhibited a logarithmic dependence on the cut. 
In Ref.~\cite{Elmfors:1997tt}, the authors computed the  rate in the real-time formalism of thermal quantum field theory (QFT),  with a  resummed photon propagator in the Hard-Thermal-Loop (HTL) approximation~\cite{Thoma:2000dc,Bellac2000},  but the imaginary part of the neutrino self-energy is time-ordered rather than retarded\footnote{It is known  that  the imaginary part of  self-energy calculated in the imaginary-time formalism of thermal QFT is equivalent to that obtained with retarded self-energy in the real-time formalism~\cite{Thoma:2000dc,Laine:2016hma}. 
In Ref.~\cite{Elmfors:1997tt}, the authors used a time-ordered amplitude as claimed in Eq.~(4.2). However, we will show in Appendix~\ref{sec:check}   that it is in fact a retarded amplitude.
}. 
Later in Ref.~\cite{Ayala:1999xn}, the chirality-flipping rate was computed with a retarded loop amplitude under the 
HTL approximation~\cite{Thoma:2000dc,Bellac2000}, 
and  a momentum-cut approach was used to separate the hard- and soft-momentum transfers~\cite{Braaten:1991dd}. These different treatments are one of the reasons that lead to the different upper bounds of $\mu_\nu$ mentioned above.

In our more elaborated computation of the chirality-flipping rate, we adopt the real-time formalism of thermal QFT and take into account the resummed photon propagator with the damping rate not limited to the HTL approximation.  
We present the master integral for the collision rate that automatically contains contributions of both photon-mediated scattering processes and plasmon decay $\gamma^*\to \bar\nu_L+\nu_R$. Since the separation of  hard- and soft-momentum contributions is known to be nontrivial~\cite{Ayala:1999xn,Besak:2012qm}, we will follow a numerical approach to  compute the master integral, which allows us to obtain the collision rate more efficiently.
Our calculation is focused on the Dirac neutrino case, but the collision rate computed in this work can be readily applied to Majorana neutrinos.

The paper is organized as follows.  
In Sec.~\ref{sec:Scomputation}, we start with a straightforward calculation of 
the collision rate from the tree-level scattering amplitude, and show that the calculation relies on the infrared cut imposed on the momentum transfer. 
A more consistent treatment requires loop calculations in the thermal QFT, which will be elaborated in Sec.~\ref{sec:Tcomputation}. 
Using the obtained collision rate, we compute the NMM correction to $N_{\rm eff}$ and derive cosmological bounds on NMM in Sec.~\ref{sec:NMM-Neff}.   
Finally, we draw our conclusions in Sec.~\ref{sec:conclusion}.

\section{$\nu_L\to \nu_R$ flipping rate from tree-level scattering amplitudes \label{sec:Scomputation}}

	

The  effective Lagrangian of a Dirac NMM is formulated as
\begin{align}\label{lag}
\mathcal{L}\supset \frac{1}{2}\mu_\nu \overline{\nu} \sigma^{\alpha \beta}\nu F_{\alpha \beta}\,,
\end{align}
where $\nu$ denotes the Dirac spinor of a neutrino, $\sigma^{\alpha\beta}=i[\gamma^\alpha,\gamma^\beta]/2$, and $F_{\alpha \beta}=\partial_{\alpha} A_{\beta}-\partial_{\beta} A_{\alpha}$ is the electromagnetic field tensor. In the chiral basis, $\nu=(\nu_L,\nu_R)^T$, we can write it as
\begin{align}\label{lag-2}
	\mathcal{L}\supset \mu_\nu \overline{\nu_L} \sigma^{\alpha \beta}\nu_R \partial_{\alpha} A_{\beta} +\rm h.c.\,,
\end{align}
which implies that the NMM operator flips the chirality of
the neutrino. Neutrinos could also possess electric dipole moments, ${\cal L}\supset\epsilon_{\nu}\overline{\nu}\sigma^{\alpha\beta}i\gamma^{5}\nu F_{\alpha\beta}/2$,
similar to Eq.~(\ref{lag}) except for an additional $i\gamma^{5}$.
Our calculations for NMM can be applied to electric dipole moments
by simply replacing $\mu_{\nu}^{2}\to\epsilon_{\nu}^{2}$ in the $\nu_{L}\to\nu_{R}$
rates because the squared amplitudes of all processes considered in
this work are not affected by the additional $i\gamma^{5}$, which 
can be seen manifestly in terms of Weyl spinors. 
 
\begin{figure}[t]
	\centering
	\includegraphics[scale=1]{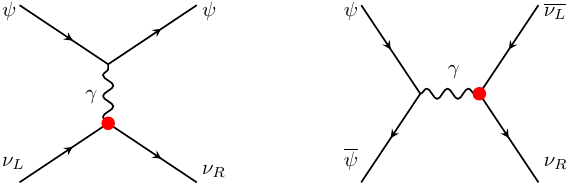}
	\caption{The $t$- and $s$-channel scattering processes for $\nu_R$ production in the early universe. The red blobs denote neutrino magnetic moments and $\psi$ denotes a generic charged fermion. 
	\label{fig:chiflip}
	}

\end{figure}

In the presence of such a chirality-flipping interaction, $\nu_R$ can be produced in the thermal bath of the early universe via the tree-level scattering processes (see Fig.~\ref{fig:chiflip}), $\psi+\overline{\psi} \to \overline{\nu_L}+\nu_R$ and $\nu_L+\psi\to \nu_R+\psi$, where $\psi$ denotes a generic charged fermion. In this paper, we assume that NMM are flavor universal and flavor diagonal for simplicity, which implies that each NMM is responsible for the thermal production of a single species of $\nu_R$.

Let us first calculate 
 the chirality-flipping rate for the tree-level scattering processes in the zero-temperature limit and then show the necessity of considering finite-temperature corrections.  
For the $s$-channel process $\psi+\overline{\psi} \to \overline{\nu_L}+\nu_R$,
 the squared amplitude reads
\begin{align}
|\mathcal{\overline{M}}|_{2\psi\to2\nu}^{2}=8\pi\alpha\mu_{\nu}^{2}\frac{tu}{s}\, ,
\end{align}
where $\alpha\equiv e^2/(4\pi)$ is the fine-structure constant,  $s,t,u$ are the Mandelstam variables, and all the fermion masses are neglected. The cross section is found to be constant in energy,
\begin{align}
	\sigma_{2\psi\to 2\nu}=\frac{\alpha \mu^2_{\nu}}{12}\,.
\end{align}
The cross section is to be used in the Boltzmann equation, 
\begin{align}
	\frac{d}{dt}n_{\nu_R}+3Hn_{\nu_R}=C_{\nu_R}^{\rm (gain)}-C_{\nu_R}^{\rm (loss)}\,,\label{eq:n-boltz}
\end{align}
where $n_{\nu_R}$ denotes the number density of $\nu_R$, $H$ is the Hubble parameter, and $C_{\nu_R}^{\rm (gain)}$ and $C_{\nu_R}^{\rm (loss)}$ are collision terms accounting for the gain and loss of $\nu_R$ due to reactions. Since our discussions will be mainly on $C_{\nu_R}^{\rm (gain)}$, we denote $C_{\nu_R}=C_{\nu_R}^{\rm (gain)}$ for brevity.
The $s$-channel contribution to $C_{\nu_R}$, denoted by $C_{\nu_R,s}$, is computed as follows~\cite{Gondolo:1990dk}:
\begin{equation}\label{eq:CnuRs}
C_{\nu_R,s}\approx   \frac{T}{8\pi^4}\int_{0}^\infty \sigma_{2\psi\to 2\nu}\,  K_1(\sqrt{s}/T)  s^{3/2} ds
\approx \frac{\alpha \mu_\nu^2}{3\pi^4}T^6\, ,
\end{equation}
where $T$ is the temperature of the SM thermal bath and $K_1$ is the modified Bessel function of order one.
In Eq.~\eqref{eq:CnuRs} we have taken the same approximations adopted in Ref.~\cite{Gondolo:1990dk} such as  neglecting the Pauli-blocking effects and using the Boltzmann distributions.

For the purpose of studying when $\nu_R$ can be in thermal equilibrium, we  define the thermally averaged collision rate, 
\begin{align}
\langle \sigma v n \rangle \equiv \frac{C_{\nu_R}}{n^{\rm eq}_{\nu_R}}\,,
\end{align}
where $n^{\rm eq}_{\nu_R}$ is the equilibrium value of $n_{\nu_R}$. For the $s$-channel collision term, it reads
\begin{align}
\langle \sigma v n \rangle_s \approx 0.04 \alpha \mu_\nu^2 T^3.\label{eq:sigmavns}
\end{align}
With $\langle \sigma v n \rangle$ defined, the condition of $\nu_R$ in thermal equilibrium is formulated as
\begin{equation}
H\lesssim \langle \sigma v n \rangle\,.\label{eq:condition}
\end{equation}
The Hubble parameter is determined by $H\approx 1.66 \sqrt{g_{\star}(T)}T^2/M_{\rm Pl}$ where $M_{\rm Pl}\approx1.22 \times 10^{19}$~GeV is the Planck mass and $g_{\star}$ is the number of effective degrees of freedom. Neglecting the temperature dependence of $g_{\star}$, we see that $H$ is proportional to $T^2$ while $\langle \sigma v n \rangle$ is proportional to $T^3$. Therefore, at a sufficiently high temperature, Eq.~\eqref{eq:condition} is always satisfied. By solving $H=C_{\nu_R}/n^{\rm eq}_{\nu_R}$ with respect to $T$, one can obtain the temperature of $\nu_R$ decoupling, $T_{\rm dec}$. For $T>T_{\rm dec}$, $\nu_R$ is in thermal equilibrium with the SM plasma. For $T<T_{\rm dec}$, $\nu_R$ is decoupled and its temperature can be computed using entropy conservation.



Next, let us include the $t$-channel contribution to the collision rate. The squared amplitude of $\nu_L+\psi\to \nu_R+\psi$ is given by
\begin{align}\label{eq:t-amp}
	|\mathcal{\overline{M}}|_{\nu \psi \to \nu \psi}^2=16\pi \alpha \mu^2_{\nu}\frac{s u}{t}\,,
\end{align}
which is known to have an IR divergence in the soft-scattering limit, $t\to 0$. 
In previous studies~\cite{Morgan:1981zy,Fukugita:1987uy},
an IR cut was imposed manually on the momentum transfer. This is qualitatively correct 
since  the photon at finite temperatures has modified dispersion relations and acquires an effective thermal mass $m_\gamma\sim e T$ in the relativistic QED plasma~\cite{Thoma:2000dc,Bellac2000}.   
Keeping   the  photon thermal  mass as an IR regulator,  
 the cross section reads
\begin{align}\label{eq:t-cs}
\sigma_{\nu \psi\to \nu \psi}= \alpha \mu_\nu^2\left(\frac{(2m_\gamma^2+s)}{s}\ln \left(\frac{m_\gamma^2+s}{m_\gamma^2}\right)-2\right).
\end{align}
Using $m_\gamma= e T/\sqrt{6}$ to be derived in Sec.~\ref{sec:Tcomputation},  we obtain the thermally averaged collision rate:
\begin{align}\label{eq:sigmavnt}
\langle \sigma v n\rangle_t \approx 
2.22
\alpha \mu_\nu^2 T^3\,,
\end{align}
where both $\psi+\nu_L \to \nu_R+\psi$ and $\bar\psi+\nu_L \to\bar\psi+ \nu_R$ have been taken into account.
By comparing Eq.~\eqref{eq:sigmavnt} to Eq.~\eqref{eq:sigmavns}, one can see that $\langle \sigma vn \rangle_t/\langle \sigma v n\rangle_s \approx  55$, which implies that the $t$-channel scattering dominates the $\nu_R$ production, at least according to the above calculation with the simple IR regulator.  

A precise computation of $\nu_R$ production rate that can  consistently  remove the IR divergence involves thermal QFT, as we will present in the next section.


\section{$\nu_L\to \nu_R$ flip rate at finite temperatures}\label{sec:Tcomputation}
\subsection{Computation method}
The IR divergence in the $t$ channel
can be canceled by taking the finite-temperature effects into account. Previously, a momentum-cut approach was introduced to separate the hard- and soft-momentum contributions~\cite{Braaten:1991dd}. The former is  calculated in the tree-level scattering amplitude with a vacuum photon propagator, while the latter takes into account the resummed photon propagator at finite temperatures. Although it has been shown that the dependence on the momentum cut would be canceled after combining the hard- and soft-momentum contributions,  this approach is based on some particular sum rules of the thermal propagators~\cite{Ayala:1999xn,Besak:2012qm}. Sometimes finding the rules is a nontrivial task. In particular, the analytic extraction of the momentum-cut dependence in the soft regime is not so simple as that in the hard domain. If we are only concerned with the total rate, a full momentum integration that automatically combines the hard and soft regimes could be more efficient, as 
applied in Ref.~\cite{Elmfors:1997tt}. Below we present the calculations in details. Readers who are not interested in the finite-temperature calculations are referred to Tab.~\ref{tab:differentrates} for the final results.


\subsection{Collision rate in the Boltzmann equation}

\begin{figure}[t]
	\centering
	\includegraphics[scale=1.5]{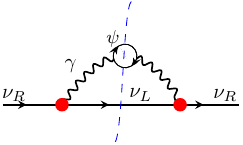}
	\caption{The self-energy diagram of $\nu_R$ in the presence of neutrino magnetic moments (red blobs). The photon propagator is resummed to include the charged-fermion loop.  In thermal QFT, the evaluation of this diagram can be used to obtain the total $\nu_R$ production rate which automatically includes the contributions of plasmon decay and  scattering processes. For instance, the dashed cut leads to a tree-level diagram corresponding to the on-shell  scattering shown in Fig.~\ref{fig:chiflip}. 
	}
	\label{fig:ChiralityFlip-1Loop}
\end{figure}

The evolution of phase-space distribution function of $\nu_R$ is governed by the  following Boltzmann equation,
\begin{align}\label{eq:f-boltz}
\left[\frac{\partial}{\partial t}-H\vec{p}\cdot\nabla_{\vec{p}}\right]f_{\nu_{R}}(\vec{p},t)=
	(1-f_{\nu_R})\Gamma_{{\nu_R}, \rm gain}-f_{\nu_R} \Gamma_{{\nu_R},\rm loss}\,,
\end{align}
where $\Gamma_{{\nu_R}, \rm gain/loss}$ denotes the gain/loss rate of  ${\nu_R}$, respectively. 
Their sum,  $\Gamma_{\nu_R,\rm tot}\equiv \Gamma_{{\nu_R},\rm gain}+\Gamma_{{\nu_R},\rm loss} $, can be physically interpreted as the rate of $f_{\nu_R}$ evolving towards equilibrium~\cite{Weldon:1983jn}\footnote{For a more explicit example, see Eq.~(3.8) in Ref.~\cite{Elmfors:1997tt}. }. 
At finite temperatures, it is related to the imaginary part of the retarded ${\nu_R}$ self-energy as follows~\cite{Weldon:1983jn} (see also Appendix~\ref{sec:weldon} for further details):
\begin{align}\label{eq:totrate}
	\Gamma_{\nu_R, \rm tot}=	\Gamma_{{\nu_R},\rm gain}+\Gamma_{{\nu_R},\rm loss}=-\frac{\text{Tr}[\slashed{p}\text{Im}\Sigma_{R}(p_{\mu})]}{E_p}\,,
\end{align}
where  $\Sigma_{R}(p_{\mu})$ denotes the retarded self-energy of $\nu_R$ and  $p_\mu=(E_p,\ \vec p)$ is the neutrino four-momentum. In the real-time formalism of thermal QFT,  the main task is to calculate the imaginary part of $\Sigma_{R}$. The diagram of  $\Sigma_{R}$, as shown in Fig.~\ref{fig:ChiralityFlip-1Loop}, consists of a $\nu_L$ propagator and a resummed $\gamma$ propagator which includes the contribution of charged-fermion loops. The tree-level scattering amplitudes in Fig.~\ref{fig:chiflip} correspond to a half of the loop diagram after cutting it symmetrically along the blue dashed line in Fig.~\ref{fig:ChiralityFlip-1Loop}. 
According to the optical theorem, the squared amplitude of the tree-level diagram can be computed from the imaginary part of the loop diagram. 
Since in the resummed photon propagator an infinite number of one-particle-irreducible (1PI) loops are actually included, one can also add another 1PI loop to the photon propagator in Fig.~\ref{fig:ChiralityFlip-1Loop} and then cut it symmetrically. 
This corresponds to the contribution of plasmon decay. 

The collision term $C_{\nu_R}$ for $\nu_R$ production previously introduced in  Eq.~\eqref{eq:n-boltz} 
is related to $\Gamma_{{\nu_R}, \rm gain}$ via 
\begin{align}\label{eq:collisionrate}
C_{\nu_R}&= \int \frac{d^3 p_{}}{(2\pi)^3} (1-f_{\nu_R})\Gamma_{\nu_R, \rm gain}
\nonumber\\[0.2cm]
&= -\int \frac{d^3 p_{}}{ (2\pi)^3 E_p}  f_{\nu_R}^{\rm eq}(1-f_{\nu_R})
\text{Tr}[\slashed{p}\text{Im}\Sigma_{R}(p_{\mu})]\,,
\end{align}
where $f^{\rm eq}_{\nu_R}$
is the distribution function in thermal equilibrium, and  we have used the unitary condition $\Gamma_{\nu_R, \rm gain}=f^{\rm eq}_{\nu_R}\Gamma_{\nu_R, \rm tot}$~\cite{Weldon:1983jn}.

\begin{figure}[t]
	\centering
	\includegraphics[scale=1.5]{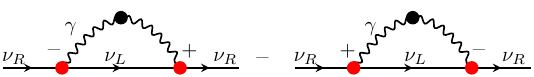}
	\caption{Two contributions, $\Sigma_{-+}(p)$ and $\Sigma_{+-}(p)$,  to the imaginary part of the retarded right-handed neutrino   self-energy. The black blob denotes the  charged-fermion loop in the resummed photon propagator, as shown in Fig.~\ref{fig:ChiralityFlip-1Loop}.
	}
	\label{fg:ImPiRnu}
\end{figure}

\subsection{Neutrino self-energy from NMM interaction}
Let us now compute the retarded  self-energy of right-handed neutrino $\Sigma_{R}(p)$. 
The imaginary part of retarded self-energy can be evaluated by\footnote{Throughout this paper, we adopt   the Keldysh basis of the real-time formalism and follow the convention of Ref.~\cite{Thoma:2000dc}.} 
\begin{equation}
\text{Im}\Sigma_{R}(p)=\frac{i}{2}\left[\Sigma_{-+}(p)-\Sigma_{+-}(p)\right],
\label{eq:imSigma}
\end{equation}
where $\Sigma_{-+}(p)$ and $\Sigma_{+-}(p)$ represent the two amplitudes in Fig.~\ref{fg:ImPiRnu}. 
The explicit forms are
\begin{align}\Sigma_{-+}(p) & =-i\mu_{\nu}^{2}\int\frac{d^{4}q}{(2\pi)^{4}}q_{\alpha}q_{\beta}\sigma^{\alpha\gamma}P_{L}S_{-+}(k)\sigma^{\beta\lambda}P_{R}\tilde{G}_{+-,\gamma\lambda}(q)\, ,\label{eq:sig-1}\\[0.3cm]
\Sigma_{+-}(p) & =-i\mu_{\nu}^{2}\int\frac{d^{4}q}{(2\pi)^{4}}q_{\alpha}q_{\beta}\sigma^{\alpha\gamma}P_{L}S_{+-}(k)\sigma^{\beta\lambda}P_{R}\tilde{G}_{-+,\gamma\lambda}(q)\,,\label{eq:sig-2}
\end{align}
where $k\equiv p+q$ and $q$ denote the momentum of $\nu_L$ and the photon. $S_{\pm\mp}$ denote the free thermal propagators of the neutrino~\cite{Thoma:2000dc}:
\begin{align}
S_{-+}(k) & =-2\pi i\,\text{sign}(k_{0})[1-f_{\nu_{L}}(k_0)]\delta(k^{2})\slashed{k}=2\pi i[-\theta(k_0)+f_{\nu_L}(|k_0|)] \delta(k^2)\slashed{k}\,,\label{freeS-+}\\[0.2cm]
S_{+-}(k) & =2\pi i\,\text{sign}(k_{0})f_{\nu_{L}}(k_0)\delta(k^{2})\slashed{k}=2\pi i [-\theta(-k_0)+f_{\nu_L}(|k_0|)]\delta(k^2)\slashed{k}\, ,\label{freeS+-}
\end{align}
where $f_{\nu_L}$ is the thermal distribution function of $\nu_L$. The last parts of Eqs.~\eqref{eq:sig-1} and \eqref{eq:sig-2},  $\tilde{G}_{\pm\mp,\mu\nu}$, denote the resummed photon propagators, which will be elucidated in  Sec.~\ref{sub:photon}.


\subsection{Resummed photon propagators\label{sub:photon}}
\subsubsection{Polarization tensors}


For the retarded photon self-energy amplitude, $\Pi_{R,\mu\nu}$, the most general tensor structure obeying the Ward identity ($q^\mu \Pi_{R,\mu\nu}=0$)
 is given by\footnote{The convention for loop amplitude is taken as  $-i\Pi_{}$  such that  the resummed propagator $\tilde{G}$ in the Dyson-Schwinger equation is written as $\tilde{G}^{-1}= G_{}^{-1}-\Pi_{}$.}
\begin{align}\label{eq:PiRdef}
	-i\Pi_{R, \mu\nu}=(-i \Pi^L_R) (P_{\mu\nu}^L)+(-i\Pi^T_R)(P_{\mu\nu}^T)\,,
\end{align}
where the longitudinal and transverse polarization tensors are given by\footnote{Here, we absorb the minus sign of $P^{L,T}$ in Weldon's convention into the definition of $\Pi^{L,T}$. }~\cite{Weldon:1982aq}
\begin{align}\label{covariantPLT}
	P^L_{\mu\nu}&=\frac{1}{q^2 |\vec q|^2}[|\vec q|^2 u_\mu +\omega(q_\mu-\omega u_\mu)][|\vec q|^2 u_\nu+\omega(q_\nu-\omega u_\nu)]\,,
	\\[0.3cm]
	P^T_{\mu\nu}&=-\eta_{\mu\nu}+u_\mu u_\nu -\frac{1}{|\vec q|^2}(q_\mu-\omega u_\mu)(q_\nu-\omega u_\nu)\,,
\end{align}
with  $q^2\equiv q_0^2-|\vec q|^2$, $\omega\equiv q^\mu u_\mu$, $\eta_{\mu\nu}=(1,-1,-1,-1)$ and $u_\mu$ the 4-velocity of the plasma. In the rest  frame,  $u_\mu=(1,\vec 0)$ and the polarization tensors  reduce  to 
\begin{align}
	P^L_{00}&=\frac{|\vec q|^2}{q^2}\,,~ P_{0i}^L=P_{00}^T=P^T_{0i}=0\,,~P_{ij}^L=\frac{q_0^2 q_i q_j}{q^2 |\vec q|^2}\,, ~P^T_{ij}=-\eta_{ij}-\frac{q_i q_j}{|\vec q|^2}\,.
\end{align}

The   retarded propagator for free photon is given by
\begin{align}
	G_{R,\mu\nu}(q)&=G_{++,\mu\nu}(q)-G_{+-,\mu\nu}(q)=\left(-\eta_{\mu\nu}+\frac{q_\mu q_\nu}{q^2}\right)\frac{1}{q^2+i\text{sign}(q_0)\epsilon}\, ,
\end{align}
where the free $2\times 2$ thermal propagator matrix elements, $G_{\pm\pm,\mu\nu}(q)$, are given by
\begin{align}
	G_{++,\mu\nu} (q)&=\left(-\eta_{\mu\nu}+\frac{q_\mu q_\nu}{q^2}\right)\left(\frac{1}{q^2+i \epsilon}-2\pi i f_\gamma(|q_0|)\delta(q^2)\right),
 \label{freeG++}
 \\[0.2cm]
G_{+-,\mu\nu}(q)&=-\left(-\eta_{\mu\nu}+\frac{q_\mu q_\nu}{q^2}\right)2\pi i \text{sign}(q_0)f_\gamma(q_0) \delta(q^2)\label{freeG+-}
\nonumber\\[0.2cm]
&=-\left(-\eta_{\mu\nu}+\frac{q_\mu q_\nu}{q^2}\right)2\pi i [\theta(-q_0)+f_\gamma(|q_0|)]\delta(q^2)\, ,
 \\[0.2cm]
G_{-+,\mu\nu}(q)&=-\left(-\eta_{\mu\nu}+\frac{q_\mu q_\nu}{q^2}\right)2\pi i \text{sign}(q_0)[1+f_\gamma(q_0)] \delta(q^2)
\nonumber \\[0.2cm]
&=-\left(-\eta_{\mu\nu}+\frac{q_\mu q_\nu}{q^2}\right)2\pi i  [\theta(q_0)+f_\gamma(|q_0|)]\delta(q^2)\, ,\label{freeG-+}
\end{align}
and $G_{--,\mu\nu} (q)=-G_{++,\mu\nu}^* (q)$. 
Then, the Dyson-Schwinger equation in terms of $\Pi^{L,T}_R$   reads
\begin{align}
	\tilde{G}_{R,\mu\nu}&=G_{R,\mu\nu}+G_{R,\mu\alpha} \Pi_{R}^{\alpha\beta} G_{R,\beta\nu} +(\Pi G)^2+...
	\nonumber \\[0.2cm]
	&=(G_R^L+G_R^L \Pi^L_R G_R^L+...)P^L_{\mu\nu}+(G_R^T+G_R^T \Pi^T_R G_R^T+...)P^T_{\mu\nu}
	\nonumber \\[0.2cm]
	&\equiv \tilde{G}_R^L P_{\mu\nu}^L+\tilde{G}_R^T P_{\mu\nu}^T\, ,
\end{align}
where we have used the orthogonality, 
\begin{align}
	P^{A}_{\mu\alpha}P^{A\alpha \nu}&=P^{A,\nu}_{\mu}\, ,  \qquad P^{L}_{\mu\alpha}P^{T,\alpha\nu}=0\, ,
\end{align}
for $A=L,T$,
and 
\begin{align}
	\tilde{G}_R^{L}(q)=\frac{1}{q^2-\Pi^L_R+i \text{sign}(q_0) \epsilon}\,, \qquad \tilde{G}_R^{T}(q)=\frac{1}{q^2-\Pi^T_R+i \text{sign}(q_0) \epsilon}\,.
\end{align}
The modifications to the photon dispersion relation arising from the longitudinal and transverse parts are now encoded  in the different   thermal scalar functions $\Pi^{L,T}_R(q)$. 

Given the free and resummed retarded propagators, we can  compute any  thermal components of  resummed $\tilde{G}_{AB, \mu\nu}$ for $A,B=\pm$ via the diagonalization approach~\cite{Aurenche:1991hi}. Explicitly, we have
\begin{align}\label{eq:diagonalization}
	\tilde{G}_{AB,\mu\nu}=U_{AC}\hat{G}_{C, \mu\nu}V_{CB}\, , \qquad 	 \Pi_{AB,\mu\nu}=V^{-1}_{AC}\hat{\Pi}_{C,\mu\nu}U^{-1}_{CB}\, ,
\end{align}
where the diagonalization matrices $U,V$ are given by~\cite{Aurenche:1991hi}
\begin{align}\label{eq:UVmatrices}
	U=f_{B/F}(q_0) e^{q_0/T}\left(	\begin{array}{cc}
		1/b_q \,   &   \pm e^{-q_0/T}/c_q \\[0.2cm]
		1/b_q   \, &   1/c_k\\
	\end{array}
	\right)\,, \qquad V= \left( \begin{array}{cc}
		b_q \,   &  \pm b_q e^{-q_0/T}\\[0.2cm]
		-c_q   \, &   -c_q\\
	\end{array}
	\right)\,,
\end{align}
where $f_{B/F}$ are distribution functions for bosons and fermions respectively.
Note that the physical result is independent of  the unspecified scalar functions $b_q, c_q$.
The diagonal matrices $\hat{G}_{\mu\nu},\hat{\Pi}_{\mu\nu}$ are constructed  by retarded/advanced propagators $G_{R/A}$ and loop amplitudes $\Pi_{R/A}$,
\begin{align}
	\hat{G}_{\mu\nu}=\left(
	\begin{array}{cc}
		G_{R,\mu\nu} \,   &   0 \\[0.2cm]
		0   \, &    G_{A,\mu\nu}\\
	\end{array}
	\right),\qquad 	\hat{\Pi}_{\mu\nu}=\left(
	\begin{array}{cc}
		\Pi_{R,\mu\nu} \,   &   0 \\[0.2cm]
		0   \, &    \Pi_{A,\mu\nu}\\
	\end{array}
	\right),
\end{align}
with $G_{A,\mu\nu}=G^*_{R,\mu\nu}$ and $\Pi_{A,\mu\nu}=\Pi_{R,\mu\nu}^*$.
We can then obtain  $\tilde{G}_{+-,\mu\nu}$, $\tilde{G}_{-+, \mu\nu}$ in  the retarded self-energy amplitude of $\nu_R$ as follows:
\begin{align}
\tilde{G}_{+-,\mu\nu}(q) & =2\pi i\,f_{\gamma}(q_0)\left[\rho^{L}(q)P_{\mu\nu}^{L}+\rho^{T}(q)P_{\mu\nu}^{T}\right],\label{resummedG+-}\\[0.2cm]
\tilde{G}_{-+,\mu\nu}(q) & =2\pi i\,\left[1+f_{\gamma}(q_0)\right]\left[\rho^{L}(q)P_{\mu\nu}^{L}+\rho^{T}(q)P_{\mu\nu}^{T}\right],\label{resummedG-+}
\end{align}
 where the 
 spectral densities $\rho^{L,T}(q)$  are given by
\begin{align}\label{eq:spectraldensity}
	\rho^{L,T}(q)= \frac{1}{\pi}\frac{\text{Im}\Pi_{R}^{L,T}(q)-\text{sign}(q_0)\epsilon}{[q^2-\text{Re}\Pi_R^{L,T}(q)]^2+[\text{Im}\Pi_R^{L,T}(q)-\text{sign}(q_0)\epsilon]^2}\, .
\end{align}
In the free limit, $\Pi_R^{L,T}=0$, $\tilde{G}_{+-,\mu\nu}$ and $\tilde{G}_{-+, \mu\nu}$   reduce  to the form given in Eqs.~\eqref{freeG+-}-\eqref{freeG-+}.

\subsubsection{Self-energy at high temperatures}
\begin{figure}[t]
	\centering
	\includegraphics[scale=1.5]{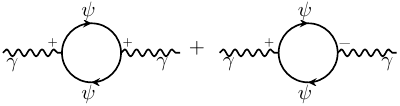}
	\caption{The  retarded    self-energy diagram of the photon from charged-fermion loop.
	}
	\label{fg:PiRgamma}
\end{figure}

The remaining task towards determining the resummed photon propagator  
is to compute  the transverse and longitudinal functions  $\Pi_R^{L,T}(q)$ from the retarded photon self-energy diagram, as shown in Fig.~\ref{fg:PiRgamma}. For a given $\psi$ in the photon self-energy loop,  the amplitude is given by
\begin{align}\label{eq:PiRgamma}
	\Pi_{R}^{\mu\nu}(q)&=-i e^2 \int \frac{d^4 k^\prime}{(2\pi)^4}\text{tr}[\gamma^\mu S_{++}(q+k^\prime)\gamma^\nu S_{++}(k^\prime)-\gamma^\mu S_{+-}(q+k^\prime)\gamma^\nu S_{-+}(k^\prime)]
	\nonumber \\[0.2cm]
	&=16\pi e^2 \int \frac{d^4 k^\prime}{(2\pi)^4}f_\psi(|k^\prime_0|)\frac{-\eta^{\mu\nu}q \cdot  k^\prime-\eta^{\mu\nu}k^{\prime 2}+q^\nu k^{\prime \mu}+q^\mu k^{\prime \nu}+2k^{\prime \nu} k^{\prime \mu}}{(q+k^{\prime})^2+i \text{sign}(q_0+k^\prime_0)\epsilon}\delta(k^{\prime 2})\, ,
\end{align}
where $S_{-+}$ and $S_{+-}$ are  given in Eqs.~\eqref{freeS-+} and \eqref{freeS+-} while the time-ordered propagator neglecting the mass of $\psi$ is given by
\begin{align}
S_{++}(k^\prime)=\left(\frac{1}{k^{\prime 2}+i \epsilon}+2\pi i f_\psi (| k^\prime_0 |)\delta(k^{\prime 2})\right) \slashed{k}^\prime \, .
\end{align}
Note that  in Eq.~\eqref{eq:PiRgamma}, we have kept only the finite-temperature correction  from  the  $f_\psi(|k_0^\prime|)$ part. 
The imaginary part arises when the charged fermions     in the loop go on-shell, which generates a Dirac $\delta$-function $\delta[(q+k^\prime)^2]$. Integrating the angle via  $\delta[(q+k^\prime)^2]$  in Eq.~\eqref{eq:PiRgamma}  would give two different regions for $|\vec{k}^\prime|$ through
\begin{align}\label{cos}
	\cos\theta_1=\frac{q^2+2q_0 |\vec{k}^\prime|}{2 |\vec{q}| |\vec{k}^\prime| }\, ,\qquad 	\cos\theta_2=\frac{q^2-2q_0 |\vec{k}^\prime| }{2  |\vec{q}||\vec{k}^\prime|}\, .
\end{align}
For $q^2<0$, it leads to 
\begin{align}\label{angle31}
	\cos\theta_1&:\quad  \frac{ |\vec{q}|-q_0}{2}< |\vec{k}^\prime|<\infty\, , \quad \text{sign}(q_0+ |\vec{k}^\prime|)=1\,, 
	 \\[0.2cm]
	\cos\theta_2&: \quad \frac{ |\vec{q}|+q_0}{2}<|\vec{k}^\prime|<\infty\, , \quad \text{sign}(q_0- |\vec{k}^\prime|)=-1\, ,\label{angle32}
\end{align}
while for $q^2\geqslant 0$, it gives
\begin{align}\label{angle4}
	\cos\theta_1&:\quad  -\frac{|\vec{q}|+q_0}{2}< |\vec{k}^\prime|<\frac{|\vec{q}|-q_0}{2}\,, \quad q_0<-|\vec{q}|\, ,\quad \text{sign}(q_0+ |\vec{k}^\prime|)=-1\,, 
	\nonumber\\[0.2cm]
	\cos\theta_2&: \quad  \frac{q_0-|\vec{q}|}{2}< |\vec{k}^\prime|<\frac{|\vec{q}|+q_0}{2}\,, \quad q_0>|\vec{q}|\,,\quad \text{sign}(q_0- |\vec{k}^\prime|)=1\,.
\end{align}
We can also see from Eq.~\eqref{cos} that if we  neglect the $q^2$ term, the conditions $|\cos\theta_{1,2}|\leqslant 1$  are only supported by  spacelike photon propagation, i.e., $-|\vec{q}|<q_0<|\vec{q}|$. This observation is equivalent to the results under the   HTL approximation, where only the corrections from the $q^2<0$ regime are considered in the resummed photon propagator.  Evaluating  the longitudinal part $\text{Im}\Pi_R^L=q^2 \text{Im}\Pi_R^{ 00}/|\vec{q}|^2$ in the
$q^2<0$ and $q^2\geqslant 0$ respectively with the full quantum statistics, we obtain
\begin{align}\label{ImPiR_L-1}
	\text{Im}\Pi_R^L\Big|_{-x_q<x_0<x_q}&=\frac{e^2 T^2}{\pi} \left(1-\frac{x_0^2}{x_q^{2}}\right)\Big(\text{Li}_2[-e^{-\frac{x_0+x_q}{2}}]-\text{Li}_2[-e^{\frac{x_0-x_q}{2}}]
	\nonumber \\[0.2cm]
	&+\frac{2}{x_q}\text{Li}_3[-e^{-\frac{x_0+x_q}{2}}]-\frac{2}{x_q}\text{Li}_3[-e^{\frac{x_0-x_q}{2}}]\Big)\, ,
	\\[0.2cm]
		\text{Im}\Pi_R^L\Big|_{x_0<-x_q}&=-\frac{e^2 T^2}{\pi}  \left(1-\frac{x_0^2}{x_q^{2}}\right)\Big(\text{Li}_2[-e^{\frac{x_0+x_q}{2}}]+\text{Li}_2[-e^{\frac{x_0-x_q}{2}}]
		\nonumber \\[0.2cm]
		&-\frac{2}{x_q}\text{Li}_3[-e^{\frac{x_0+x_q}{2}}]+\frac{2}{x_q}\text{Li}_3[-e^{\frac{x_0-x_q}{2}}]\Big)\, ,\label{ImPiR_L-2}
\end{align}
where $x_0\equiv q_0/T, x_q\equiv |\vec{q}|/T$, and  $\text{Li}_n$ is the  polylogarithm function of order $n$. 
Similarly, the transverse part $\text{Im}\Pi_{R}^T= (\delta_{ij}-q_i q_j/|\vec k|^2)\text{Im}\Pi_{R}^{ij}/2$ is given by
\begin{align}\label{ImPiRT-1}
	\text{Im}\Pi_R^T\Big|_{-x_q<x_0<x_q}&=\frac{e^2 T^2}{\pi} \left(1-\frac{x_0^2}{x_q^{2}}\right)\Big(\frac{1}{2}\text{Li}_2[-e^{-\frac{x_0-x_q}{2}}]-\frac{1}{2}\text{Li}_2[-e^{-\frac{x_0+x_q}{2}}]+\frac{1}{x_q}\text{Li}_3[-e^{\frac{x_0-x_q}{2}}]
	\nonumber \\[0.2cm]
&-\frac{1}{x_q}\text{Li}_3[-e^{-\frac{x_0+x_q}{2}}]-\frac{x_q}{8}\Big[2\ln\Big(\frac{e^{x_0/2}+e^{x_q/2}}{e^{(x_0+x_q)/2}+1}\Big)+x_0\Big]
\Big)\,,
\\[0.2cm]
\text{Im}\Pi_R^T\Big|_{x_0<-x_q}&=\frac{e^2 T^2}{\pi} \left(1-\frac{x_0^2}{x_q^{2}}\right)\Big(\frac{1}{2}\text{Li}_2[-e^{-\frac{x_0-x_q}{2}}]+\frac{1}{2}\text{Li}_2[-e^{\frac{x_0+x_q}{2}}]+\frac{1}{x_q}\text{Li}_3[-e^{\frac{x_0-x_q}{2}}]
\nonumber \\[0.2cm]
&-\frac{1}{x_q}\text{Li}_3[-e^{\frac{x_0+x_q}{2}}]-\frac{x_q}{8}\Big[2\ln\Big(\frac{e^{x_0/2}+e^{x_q/2}}{e^{(x_0+x_q)/2}+1}\Big)-x_q\Big]
\Big)\, .\label{ImPiRT-2}
\end{align}
In the above equations, we have dropped the contributions of $\text{Im}\Pi_R^{L,T}$ in the  $x_0>x_q$ region. As will be found out in Sec.~\ref{sec:fullrate}, this region is not supported by the integration of photon momentum $q$.

It should be pointed out that  the above results for $\text{Im}\Pi_R^{L,T}$  are exact without the HTL approximation. If we go to the limit $(|\vec{q}|\pm q_0)/2=0$ in Eqs.~\eqref{angle31}-\eqref{angle32}  and take the Fermi-Dirac statistics,  $	\text{Im}\Pi_R^{L,T}$ at $q^2<0$ would reduce to the known results in the HTL approximation~\cite{Thoma:2000dc}, i.e.,
\begin{align}\label{eq:ImHTL}
	\text{Im}\Pi_R^L=-\frac{\pi e^2 T^2}{6}\frac{q^2 q_0}{|\vec q|^3}\, ,
\quad 	\text{Im}\Pi_R^T=\frac{\pi e^2 T^2}{12}\frac{q^2q_0}{|\vec q|^3}\,.
\end{align}
Since the imaginary parts can be analytically integrated even with the full quantum statistics due to the presence of two Dirac $\delta$-functions, we will use $\text{Im}\Pi_R^{L,T}$ without the HTL approximation. 
Taking further into account the contributions in the  $q^2 \geqslant 0$ region, the contributions of scattering with both hard- and soft-momentum transfers will be included simultaneously. 
On the other hand, an analytic integration with the full quantum statistics cannot be obtained for the real part of $\Pi_R^{L,T}$. Given that  the real part 	$\text{Re}\Pi^{L,T}_R(q)\sim \alpha T^2$   plays the role of IR regulator in the soft-momentum transfer $q^2\ll T^2$ but only   serves as sub-leading corrections to the  photon spectral density   in the  hard-momentum transfer $q^2\gg \alpha_{} T^2$, 
we can  apply the HTL-approximated  results~\cite{Carrington:1997sq,Thoma:2000dc}:
\begin{align}
	\text{Re}\Pi^L_R(q)&=-\frac{2m_\gamma^2q^2}{|\vec{q}|^2}\left(1-\frac{q_0}{2|\vec{q}|} \ln\left|\frac{q_0+|\vec{q}|}{q_0-|\vec{q}|}\right|\right),
	\\[0.2cm]
	\text{Re}\Pi_R^T(q)&=\frac{m_\gamma^2q_0^2}{|\vec{q}|^2}\left(1-\frac{q^2}{2|\vec{q}|q_0}\ln\left|\frac{q_0+|\vec{q}|}{q_0-|\vec{q}|}\right|\right),
\end{align}
for the photon spectral density $\rho^{L,T}(q)$ in the full momentum space.

The computation thus far considers  only the contribution of the electron.
At temperatures above the QCD phase transition, one needs to include  quarks into the charged-fermion loop. 
	In general, for $\nu_R$ decoupling above a few hundred GeV, all the SM charged fermions and the charged $W$ boson can contribute to the photon dispersion relation in Fig.~\ref{fig:ChiralityFlip-1Loop}. 
Their contributions can be taken into account 
by  replacing
\begin{align}\label{eq:enhancefactor}
e^{2}\to c_{\psi}e^{2}\,,\ \ c_{\psi}\equiv \sum_{\psi}Q_{\psi}^{2}\,,
\end{align} 
where  $Q_{\psi}$ denotes the  electric charge of $\psi$.  
Here the summation goes over all charged particles that are relativistic in the thermal bath.  All quarks at a sufficiently high temperature contribute a combined factor of $\left(1/3\right)^{2}\times3\times3+\left(2/3\right)^{2}\times3\times3=5$ and all charge leptons contribute a factor of $3$. For the $W$ boson, we assume its contribution is the same as the electron, though strictly speaking it would require a more dedicated calculation. When the temperature is not sufficiently high, we treat $c_{\psi}$ as a step function of $T$, including only charged particles with masses below $T$.

\subsection{Full result of the  collision rate}\label{sec:fullrate}
Combining results in the above calculations, we can write the imaginary retarded amplitude as
\begin{align}\label{eq:ImPiRnu-1}
	\text{Im}\Sigma_{R}(p) & =\frac{\mu_{\nu}^{2}}{2(2\pi)^{2}}\int d^{4}qq_{\alpha}q_{\beta}\left[f_{\gamma}(q_0)+f_{\nu_{L}}(k_0)\right]{\rm sign}(k_{0})\delta(k^{2})\nonumber \\[0.2cm]
	& \times\left[\rho^{L}(q)P_{\gamma\lambda}^{L}+\rho^{T}(q)P_{\gamma\lambda}^{T}\right]\left[\sigma^{\alpha\gamma}P_{L}\slashed{k}\sigma^{\beta\lambda}P_{R}\right]\,.
\end{align}
Note that due to the presence of $\delta(k^{2})$,  which dictates $k^{2}=(p+q)^{2}=2(q_{0}p_0-\vec{p}\cdot\vec{q})+q^{2}=0$, we obtain
\begin{align}
	-2|\vec{q}||\vec{p}|-q^{2}\leqslant2q_{0}p_0\leqslant2|\vec{q}||\vec{p}|-q^{2}\thinspace.\label{eq:inequality}
\end{align}
For $q^{2}>0$ (i.e. $q_{0}>|\vec{q}|$ or $q_{0}<-|\vec{q}|$), the
second inequality in Eq.~\eqref{eq:inequality} would not be satisfied
if we take the positive branch $q_{0}>|\vec{q}|>0$.  In this case, we find that $k_{0}$
is always negative. Similarly, for $q^{2}<0$ we find that $k_{0}$ is
always positive. Hence the sign function takes
\begin{equation}
	{\rm sign}(k_{0})=\begin{cases}
		-1 & (\text{for }q^{2}>0)\\
		1 & (\text{for }q^{2}<0)
	\end{cases}\thinspace.\label{eq:angle}
\end{equation}

Assembling all the pieces in the previous calculation, the trace in Eq.~\eqref{eq:collisionrate} is given by
\begin{align}
&	\text{tr}[\slashed{p}\text{Im}\Sigma_R]
	= -\frac{\mu_\nu^2}{8\pi^2|\vec p|} \int dq_0 |\vec q| d|\vec{q}| [f_\gamma (q_0)+f_{\nu_L}(p_0+q_0)]\Big[\Theta_1(q_0,|\vec{q}|,|\vec{p}|)-\Theta_2(q_0,|\vec{q}|,|\vec{p}|)\Big]
	\nonumber\\[0.2cm]
	&\times \left[-|\vec q|^2\left(1-\frac{q_0^2}{|\vec q|^2}\right)^2(q_0+2p_0)^2\frac{\text{Im}\Pi_R^L(q_0,|\vec q|)}{[q^2-\text{Re}\Pi_R^L(q_0,|\vec q|)]^2+[\text{Im}\Pi_R^L(q_0,|\vec q|)]^2} \right.
	\nonumber \\[0.2cm]
	& \left. +|\vec q|^2\left(1-\frac{q_0^2}{|\vec q|^2}\right)^2[(q_0+2p_0)^2-|\vec q|^2]\frac{\text{Im}\Pi_R^T(q_0,|\vec q|)}{[q^2-\text{Re}\Pi_R^T(q_0,|\vec q|)]^2+[\text{Im}\Pi_R^T(q_0,|\vec q|)]^2}\right]\,, 
\label{trace-sum}
\end{align}
where  the Heaviside functions are defined as
\begin{align}\label{eq:Theta1}
	\Theta_1&\equiv \theta(-q_0-|\vec q|)\theta[|\vec q|^2-(2p_0+q_0)^2]\, ,
	\\[0.2cm]
	\Theta_2&\equiv \theta(|\vec q|^2-q_0^2)\theta[ 2p_0+q_0-|\vec q|]\, .\label{eq:Theta2}
\end{align}
Finally, we obtain the master integral for the full collision rate as follows:
\begin{align}\label{eq:masterint}
	\langle \sigma vn\rangle_{\rm full} &=\frac{c_{\psi}\alpha_{} \mu_\nu^2 T^3}{3\pi^2 \zeta(3)}\int_{0}^\infty dx_p dx_0 x_q dx_q f^{\rm eq}_{\nu_R}(x_p) [1-f^{\rm eq}_{\nu_R}(x_p)] [f_\gamma (x_0)+f_{\nu_L}(x_0+x_p)]
	\nonumber \\[0.2cm]
	&\times \left(\Theta_1 -\Theta_2\right) \left[\frac{-x_q^2\left(1-x_0^2x_q^{-2}\right)^2 (x_0+2x_p)^2\text{Im}\tilde{\Pi}_R^L(x_0,x_q)}{[x_0^2-x_q^2-\text{Re}\Pi_R^L(x_0,x_q)]^2+[4c_{\psi} \alpha_{} \text{Im}\tilde{\Pi}_R^L(x_0,x_q)]^2} \right.
		\nonumber \\[0.2cm]
	&\left.+\frac{x_q^2(1-x_0^2x_q^{-2})^2[(x_0+2x_p)^2-x_q^2]\text{Im}\tilde \Pi_R^T(x_0,x_q)}{[x_0^2-x_q^2-\text{Re}\Pi_R^T(x_0,x_q)]^2+[4c_{\psi} \alpha_{} \text{Im}\tilde \Pi_R^T(x_0,x_q)]^2}\right],
\end{align}
where $(x_0,\ x_q,\ x_p)\equiv (q_0,\  |\vec{q}|,\ |\vec{p}|)/T$, $\text{Im}\tilde{\Pi}_{R}^{L,T}(x_{0},x_{q})\equiv\pi e^{-2}T^{-2}\text{Im}\Pi_{R}^{L,T}(x_{0},x_{q})$, 
 and  $c_{\psi}$ is the enhancement factor defined in Eq.~\eqref{eq:enhancefactor}.
The three-dimensional integral can be integrated numerically, which yields\footnote{Our code to generate the results below is publicly available at \url{https://github.com/Shao-Ping-Li/NMM_Neff}. } 
\begin{align}\label{eq:sigmavnfull}
	\langle \sigma v n\rangle_{\rm full} \approx 
	6.47 \alpha_{} \mu_\nu^2 T^3\,,
\end{align}
where we have used $c_{\psi}=9$ to include the contribution of all charged particles.
If only $\psi=e$ is included (i.e.~$c_{\psi}=1$), we would have $\langle \sigma v n\rangle_{\psi=e}   \approx 1.53\alpha_{} \mu_\nu^2 T^3$.
If we use the HTL-approximate results given in Eq.~\eqref{eq:ImHTL}, 
 we obtain $\langle\sigma vn\rangle\approx1.84\alpha\mu_{\nu}^{2}T^{3}$
and $\langle\Gamma_{{\rm tot}}\rangle\equiv\int\frac{d^{3}p}{(2\pi)^{3}}\Gamma_{{\rm tot}}f_{\nu_{R}}/\int\frac{d^{3}p}{(2\pi)^{3}} f_{\nu_{R}}\approx1.99\alpha\mu_{\nu}^{2}T^{3}$.
The latter should be compared with the rate $\langle\Gamma_{{\rm tot}}\rangle\approx1.81\alpha\mu_{\nu}^{2}T^{3}$
obtained in Ref.~\cite{Elmfors:1997tt}. Note that the difference
between $\langle\sigma vn\rangle$, which can be written as $\langle(1-f_{\nu_{R}})\Gamma_{{\rm tot}}\rangle$
according to the definition, and $\langle\Gamma_{{\rm tot}}\rangle$
is that the former is slightly suppressed by the Pauli-blocking factor
$(1-f_{\nu_{R}})$.
From Eq.~\eqref{eq:sigmavnt}, we can see that the result derived from the tree-level scattering amplitude with a photon thermal mass cut $m_\gamma=e T/\sqrt{6}$ is   overestimated. 

\begin{table}[t]
	\begin{center}
		\begin{tabular}{l|c}
			\hline\hline
		 	Thermally averaged $\nu_L\to \nu_R$ rate &  $ \langle \sigma v n\rangle/ \alpha \mu_\nu^2 T^3$  
		 	\\ \hline
		 Zero-T QFT + $m_\gamma$ cut + $e^{\pm}$ plasma + Boltzmann statistics    	 & 	2.26
		\\[1mm]  
			Finite-T QFT + $e^{\pm}$ plasma + HTL approximation	   	   &   1.84  	
			\\[1mm] 			
			Finite-T QFT + $e^{\pm}$ plasma + full quantum statistics	   	   &   1.53  	
			\\[1mm]  
		Finite-T QFT + $\psi\overline{\psi}$ plasma + full quantum statistics	   	&	6.47
		\\[1mm] 			
		\hline \hline
		\end{tabular}
	\end{center}\vspace{-0.3cm}
	\caption{ The thermally averaged chirality-flipping rates computed in different methods.  ``Zero-T QFT + $e^{\pm}$  plasma  + Boltzmann statistics'' denotes the calculation  in Sec.~\ref{sec:Scomputation}, where we use Feynman rules of zero-temperature QFT with a finite photon mass cut $m_\gamma=e T/\sqrt{6}$ and the Boltzmann statistics in the relativistic electron-positron ($e^{\pm}$) plasma.
In the finite-T QFT approach  from Sec.~\ref{sec:Tcomputation}, finite-temperature effects and full quantum statistics (Fermi-Dirac/Bose-Einstein) are taken into account. 
	\label{tab:differentrates}}
\end{table}

The comparison of our final results obtained in the finite-temperature approach (together with quantum statistics) to the result obtained in Sec.~\ref{sec:Scomputation} is presented in Tab.~\ref{tab:differentrates}.

\subsection{Plasmon decay}

At the end of this section, we would like to present an estimate of the plasmon decay  ($\gamma^*\to \bar\nu_L+\nu_R$) effect, which would be important when the charged fermions become non-relativistic while their number densities remain high, such as in some stellar environments~\cite{Raffelt1996}. The  contribution  of plasmon decay is already included in the master formula, Eq.~\eqref{eq:masterint}, in which it corresponds to $q^2=\text{Re}\,\Pi_R^{L,T}(q)$.


For a simple estimate,
the decay rate can   be calculated by using the $S$-matrix formalism. With the effective Lagrangian Eq.~\eqref{lag}, it is straightforward to obtain the decay width as follows:
\begin{align}\label{eq:plasmondecay}
\Gamma_{\gamma^*\to 2\nu}=\frac{\mu_\nu^2q^4}{16\pi E_{\gamma^*}},
\end{align}
where $q^2=E_{\gamma^*}^2-|\vec q|^2$. 
The thermally averaged rate can be written as
\begin{align}\label{eq:<Gamma>}
\langle \Gamma_{\gamma^*\to 2\nu}\rangle \equiv \frac{1}{n_{\nu_R}^{\rm eq}}\int\frac{d^3 |\vec q|}{(2\pi)^3 }f_{{\gamma}}(E_{\gamma^*}) \Gamma_{\gamma^*\to 2\nu}\,.
\end{align}

In the non-relativistic regime ($T\ll m_{\psi}$ ) and the HTL approximation, we obtain from Eq.~\eqref{eq:PiRgamma} the photon thermal mass which depends on the photon polarization:
\begin{align}\label{eq:PiRlowT}
\text{Re}	\Pi_R^L(q)\approx 2\frac{q^2}{q_0^2}\frac{e^2n_{\psi}}{m_{\psi}} \left(1-\frac{5T}{2m_{\psi}}\right)\, , \quad \text{Re}	\Pi_R^T(q)\approx2 \frac{e^2n_{\psi}}{m_{\psi}} \left(1-\frac{5T}{2m_{\psi}}\right)\, ,
\end{align}
where $n_{\psi}$ denotes the  number density of $\psi$. 
Here we would like to make a comparison to the results in Ref.~\cite{Braaten:1993jw} where the 
photon dispersion relation for $T\ll m_e$ has also been calculated  in the stellar plasma. 
The main difference in the stellar plasma is that, as an electrically neutral medium, its positively charged particles are protons (including protons in nuclei), whose contributions to the photon thermal masses are negligible due to the heavy proton/nucleus mass.
In the thermal plasma of the early universe, we have equally high densities of both electrons and positrons.  
Therefore, the photon thermal masses in Eq.~\eqref{eq:PiRlowT} are twice as large as the results for the stellar plasma. 

Taking the approximate dispersion relation $q^2\approx m_\gamma^2$  with $m_\gamma^2\approx 2e^2 n_{\psi}/m_{\psi}$ (i.e. only the transverse polarization mode is used) and the nonrelativistic limit of $n_{\psi}$, 
\begin{align}
n_{\psi}=e^{-m_{\psi}/T}  \frac{(m_{\psi} T)^{3/2}}{\sqrt{2}\pi^{3/2}}\,,
\end{align}
 we obtain the thermally averaged production rate
\begin{align}\label{eq:plasmondecayrate}
\langle \Gamma_{\gamma^*\to 2\nu}\rangle\approx \frac{4\alpha^2 \mu_\nu^2m_{\psi} T^2}{3\pi^2 \zeta(3)}e^{-2m_{\psi}/T}\ \ \ \  (T\ll m_{\psi})\,.
\end{align}
We can see that the rate from plasmon decay is at $\mathcal{O}(\alpha^2)$ and is exponentially suppressed  when $T\ll m_{\psi}$. 
In the relativistic regime, on the other hand,  the calculation is similar and we find
\begin{align}\label{eq:plasmondecayrate-2}
\langle \Gamma_{\gamma^*\to 2\nu}\rangle \approx 0.048\alpha^2 \mu_\nu^2 T^3\ \ \ \  (T\gg m_{\psi})\,.
\end{align}


Overall, the contribution of plasmon decay to $\nu_R$ production in the early Universe is subdominant because the results in Eqs.~\eqref{eq:plasmondecayrate} and \eqref{eq:plasmondecayrate-2} are proportional to $\alpha^2$. 

%

\section{NMM bounds from $N_{\rm eff}$ constraints}\label{sec:NMM-Neff}

With the $\nu_R$ production rate obtained (see Tab.~\ref{tab:differentrates}), we are ready to relate NMM to $N_{\rm eff}$ and use  cosmological measurements of $N_{\rm eff}$ to constrain NMM. 

\begin{figure}[t]
	\centering
	\includegraphics[scale=0.7]{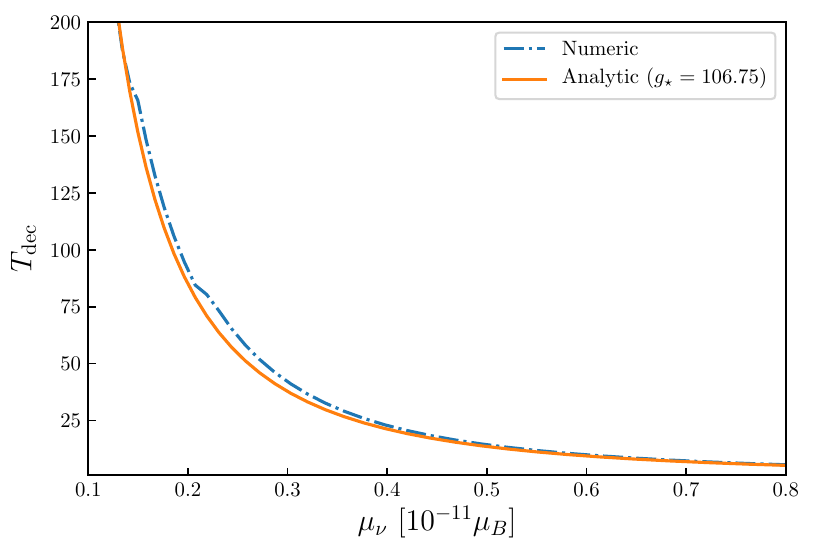}
	\caption{ The full numerical relation between the NMM $\mu_\nu$ and the decoupling temperature of right-handed neutrinos $T_{\rm dec}$. For comparison, the analytic relation in Eq.~\eqref{eq:NMM-Tdec} obtained by fixing  $g_{\star}=106.75$  and $\langle \sigma v n\rangle/T^3=6.47$ is also shown.	
	\label{fig:NMM_Tdec}
	}

\end{figure}

As we have discussed in Sec.~\ref{sec:Scomputation}, due to $\langle \sigma v n\rangle\propto T^3$ and $H\propto T^2$, we expect that $\nu_R$ is in thermal equilibrium at sufficiently high temperatures and decouples from the thermal bath at low temperatures. 
The decoupling temperature, $T_{\rm dec}$, is determined by solving $H=\langle \sigma v n\rangle$, i.e.,
\begin{equation}
\frac{1.66}{M_{{\rm Pl}}}\sqrt{g_{\star}(T_{\rm dec})}T_{{\rm dec}}^2=
\langle \sigma v n\rangle\,.\label{eq:T-dec}
\end{equation}
In general, as  $g_{\star}$ and $\langle \sigma v n\rangle/T^3$  both vary with the temperature, one has to solve  Eq.~\eqref{eq:T-dec} numerically.  Nevertheless, we can still have an approximate estimate by  setting  $g_{\star}=106.75$ and $\langle \sigma v n\rangle/T^3=6.47$, which gives
\begin{align}\label{eq:NMM-Tdec}
\mu_\nu  \approx 1.8 \times 10^{-12}\left(\frac{100\ \rm GeV}{T_{\rm dec}}\right)^{1/2}\mu_B\,.
\end{align}
The full numerical relation between  $T_{\rm dec}$ and $\mu_\nu$ is shown  in Fig.~\ref{fig:NMM_Tdec}, 
where the temperature dependence of $g_{\star}(T)$ is taken from Ref.~\cite{Wallisch:2018rzj} and $c_{\psi}$ takes the step function as mentioned below Eq.~\eqref{eq:enhancefactor}.
For comparison, the approximate relation in Eq.~\eqref{eq:NMM-Tdec} is also shown in  Fig.~\ref{fig:NMM_Tdec}.

The contribution of $\nu_R$ to the effective neutrino number excess, $\Delta N_{\rm eff}$, depends on $T_{\rm dec}$, and the relation can be derived from entropy conservation---see, e.g.,~\cite{Abazajian:2019oqj,Luo:2020sho,Li:2021okx}. Given $T_{\rm dec}$,   $\Delta N_{\rm eff}$ is determined by
\begin{align}\label{eq:Neffdef}
	\Delta N_{\rm eff} =  N_R \left[\frac{4}{43}g_{*}(T_{\rm dec})\right]^{-4/3},
\end{align}
where $N_R$ is the number of thermalized light $\nu_R$. Although $N_R=3$ is the most natural assumption because
for three Dirac neutrinos with NMM all $\nu_R$'s are expected to thermalize at a sufficiently high $T$, it is still possible to have $N_R=2$ or $1$. For instance,  it could be that not all neutrinos are Dirac so that there are only one or two light $\nu_{R}$ species present. Even if all neutrinos are Dirac, the actual temperature required by $H=\langle \sigma v n\rangle$ could be too high so that the universe has never reached such a high temperature (e.g. if the temperature is above the reheating temperature after inflation). Therefore, we keep the possibilities of $N_R=2$ and $1$ in our discussion below.

\subsection{Results}

\begin{figure}[t]
	\centering
	\includegraphics[width=0.7\textwidth]{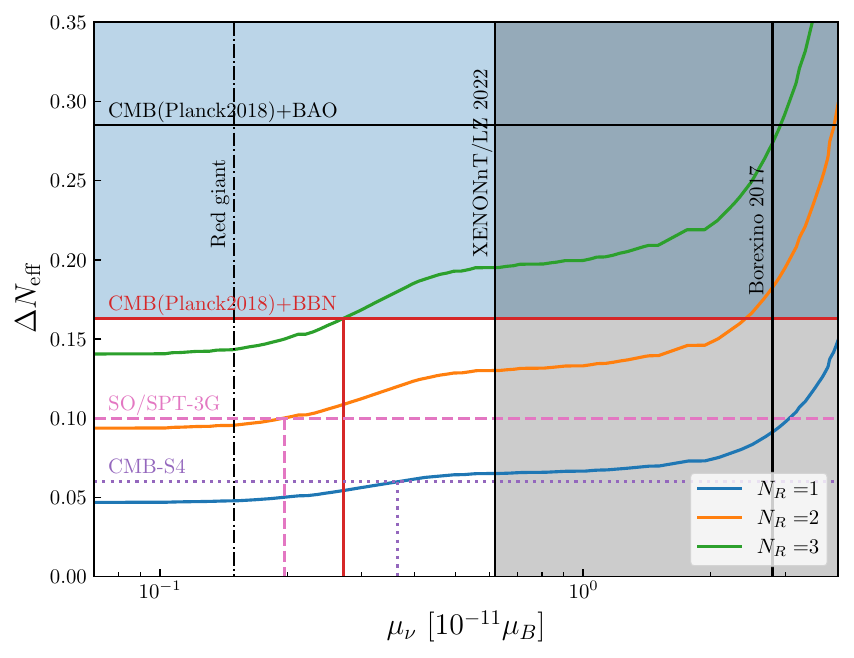}
	\caption{ 
The dependence of $\Delta N_{\rm eff}$ on $\mu_\nu$, assuming the number of $\nu_R$ thermalized via NMM is $N_R=1$, $2$, or $3$.
The red horizontal line represents the current best limit on $\Delta N_{\rm eff}$ from the combination of latest CMB (Planck 2018) and BBN data.  For $N_R=3$, it can be recast to the corresponding bound on $\mu_{\nu}$, as indicated by the red vertical line.   Future experiments such as SO/SPT-3G and CMB-S4 will be able to probe the $N_R=2$ and $N_R=1$ scenarios. 
Other constraints are explained in the text.
	}
	\label{fig:NMMNeff}
\end{figure}

Applying the numerical  $T_{\rm dec}(\mu_\nu)$ relation in Fig.~\ref{fig:NMM_Tdec} to Eq.~\eqref{eq:Neffdef}, we plot $\Delta N_{\rm eff}$ as a function of $\mu_\nu$ for  $N_{R}=1,2, 3$ in Fig.~\ref{fig:NMMNeff}, together with cosmological bounds on $\Delta N_{\rm eff}$ and other known bounds on $\mu_{\nu}$. The  bounds are explained as follows. 
 
Currently, the best measurements of $N_{\rm eff}$ come from combinations of CMB, BAO, and BBN observations. We adopt results of the CMB+BAO combination from the Planck 2018 publication~\cite{Planck:2018vyg} and the CMB+BBN combination from Ref.~\cite{Fields:2019pfx}\footnote{We take the result from the CMB+BBN column in Tab.~5 of the erratum of Ref.~\cite{Fields:2019pfx}. }:
\begin{align}
N_{{\rm eff}} & =2.99\pm0.17\thinspace,\ \ \ \ (\text{CMB+BAO})\thinspace,\label{eq:neff-1}\\
N_{{\rm eff}} & =2.830\pm0.189\thinspace,\ (\text{CMB+BBN})\thinspace.\label{eq:neff-2}
\end{align}
Subtracting the standard value, $N^{\rm st.}_{{\rm eff}}=3.045$~\cite{deSalas:2016ztq,EscuderoAbenza:2020cmq,Akita:2020szl}\footnote{Recent analyses favor   $N^{\rm st.}_{{\rm eff}}=3.044$~\cite{Gariazzo:2019gyi,Escudero:2020dfa,Froustey:2020mcq,Bennett:2020zkv}. This can not change our result significantly.}, we obtain the upper
limits $\Delta N_{{\rm eff}}<0.285$ and $\Delta N_{{\rm eff}}<0.163$
at $95\%$ (2$\sigma$) C.L., respectively. 
 


\begin{table}
	\centering
	\renewcommand{\arraystretch}{1.4}
	\begin{tabular}{l|c|c}
		\hline\hline
		Experiments/Observations & Upper Limit & Ref. \\ 
		\hline 
		XENONnT ($\nu+e^-$ scatt.) & 
		$6.3 \times 10^{-12} \mu_B $   & \cite{XENON:2022mpc}\\
		LZ ($\nu+e^-$ scatt.) & $6.2\times 10^{-12} \mu_B $ & 
		\cite{AtzoriCorona:2022jeb}\\ 
		Borexino ($\nu+e^-$ scatt.)  & $28\times 10^{-12} \mu_B $ & 
		\cite{Borexino:2017fbd}\\ 
		\hline	
		\multirow{2}{*}{Astrophysics (red giant)} & $2.2\times 10^{-12} \mu_B$ & 
		\cite{Diaz:2019kim} \\
		 & $1.5\times 10^{-12} \mu_B$& 
		\cite{Capozzi:2020cbu} \\ 
		\hline
		BBN+CMB for $N_R=3$ & $2.7\times 10^{-12} \mu_B $ & 
		This work\\
		SO for $N_R=2$ (future) & $2.0 \times 10^{-12} \mu_B $ & 
		This work\\ 
		CMB-S4 for $N_R=1$ (future) & $3.7\times 10^{-12} \mu_B $& 
		This work\\ 
		\hline	\hline
	\end{tabular}
	\caption{Severe   bounds of $\mu_\nu$ at $\mathcal{O}(10^{-12}\mu_B)$ from recent terrestrial, astrophysical and cosmological  experiments.}
	\label{tab:NMMbound}
\end{table}

Future CMB experiments such as the Simons Observatory (SO) and CMB
Stage-IV (CMB-S4) can reach $\Delta N_{{\rm eff}}<0.1$ \cite{Galitzki:2018wvp,SimonsObservatory:2019qwx}
and $\Delta N_{{\rm eff}}<0.06$ \cite{CMB-S4:2016ple,Abazajian:2019eic}
at $2\sigma$ C.L., respectively. They are plotted in Fig.~\ref{fig:NMMNeff}
as dashed lines. The South Pole Telescope (SPT-3G) \cite{SPT-3G:2014dbx}
has approximately the same sensitivity reach as SO. So we refer to
their common limit as SO/SPT-3G.

Low-energy elastic $\nu+e^-$ scattering data in neutrino and dark matter detectors can be used to constrain NMM. In Tab.~\ref{tab:NMMbound}, we list such bounds from the very recent XENONnT~\cite{XENON:2022mpc} and LUX-ZEPLIN (LZ)~\cite{LZ:2022ufs} experiments as well as that from Borexino~\cite{Borexino:2017fbd} which used to be the strongest laboratory constraint. Note that these bounds are all derived from solar neutrinos which change flavors when arriving at the Earth. At low energies, the probabilities of solar neutrinos appearing as $\nu_e$, $\nu_{\mu}$, and $\nu_{\tau}$ are around 56\%, 22\%, and 22\%, respectively~\cite{Xu:2022wcq}. Here we assume flavor-universal and flavor-diagonal NMM, for which the reported bounds in these experiments can be directly compared to our results. If one focuses on a NMM of a specific flavor, then the above probabilities should be taken into account and would lead to weaker bounds. We refer to Ref.~\cite{deGouvea:2022znk} for a more dedicated treatment of this issue.

In addition to laboratory bounds, we also include the astrophysical bounds derived from the red-giant branch,  $\mu_\nu< 2.2 \times 10^{-12}\mu_B$~\cite{Diaz:2019kim} and  $\mu_\nu< 1.5 \times 10^{-12}\mu_B$~\cite{Capozzi:2020cbu}. 
In Fig.~\ref{fig:NMMNeff}, we only plot the latter to avoid cluttering. Due to potential astrophysical uncertainties, this bound is presented as a dash-dotted line. 



As can be read from Fig.~\ref{fig:NMMNeff}, if three $\nu_R$'s are thermalized via NMM in the early Universe the current BBN+CMB data would set an upper limit on $\mu_\nu$ with
\begin{align}\label{eq:BBN+CMBbound}
	\text{BBN+CMB}:\quad  \mu_\nu<2.7\times 10^{-12}\mu_B\ \  (\text{for }N_R=3)\,. 
\end{align}
It should be noted that for $N_{R}=3$, the minimal value of $\Delta N_{\rm eff}$ is 0.14, as can be seen from the $N_{R}=3$ curve at lower $\mu_\nu$ values  in Fig.~\ref{fig:NMMNeff}. 
Future experiments like SO/SPT-3G and CMB-S4 are sensitive to  $\Delta N_{\rm eff}$ below this level. 
If $\Delta N_{\rm eff}$ is found to be smaller than 0.14 in future measurements,  the $N_R=3$  scenario  (i.e.~three $\nu_R$'s being thermalized via NMM) would be ruled out. There are multiple possibilities to go beyond this scenario: (i) the effective NMM vertex opens up at high temperatures, leading to the transition from freeze-out to freeze-in (see e.g.~\cite{Luo:2020sho,Luo:2020fdt}); (ii) as $T$ increases above the electroweak scale, $g_{\star}(T)$  continues increasing  due to new particles beyond the SM; (iii) the number of $\nu_R$ that can be possibly thermalized via NMM is less than three. 
In either (i) or (ii), the curves in Fig.~\ref{fig:NMMNeff} will further decrease  as $\mu_{\nu}$ decreases, but this part would be quite model dependent. We will investigate this in future work. As for (iii), one may consider $N_R=2$ or $1$. 
 For $N_R=2$, the SO/SPT-3G sensitivity is able to reach $\mu_\nu<2.0\times 10^{-12}\mu_B$ while for $N_R=1$ the future CMB-S4 sensitivity can set $\mu_\nu<3.7\times 10^{-12}\mu_B$. 
We see from  Fig.~\ref{fig:NMMNeff} that all these cosmological bounds are stronger than the laboratory ones from  XENONnT, LZ, and Borexino, and comparable to astrophysical ones. 
Note that the strongest astrophysical bound cuts the $N_{R}=3$ curve at $\Delta N_{{\rm eff}}=0.143$
while the curve  becomes flat at $0.141$. So there is the possibility for upcoming precision measurements of $N_{\rm eff}$ to make a discovery.  If  a small $\Delta N_{{\rm eff}}$ between $0.143$
and $0.141$ is probed, it can be attributed to NMM smaller than the strongest
astrophysical bounds. What would be more interesting is that future
experiments could measure an even smaller $\Delta N_{{\rm eff}}$ below $0.141$. In this case, it could probe the freeze-in regime
which is sensitive to how the effective vertex opens up at high energies. We leave this possibility for future exploration. 

%


\subsection{Discussions}

 We would like to make a few comments on the underlying assumptions
of our results. Large NMM arising  from new physics models   typically
involve new energy scales (denoted by $\Lambda$), as well as  new heavy particles
that directly couple to $\nu_{R}$. At temperatures around or above
$\Lambda$, the effective NMM vertex might be invalid in the calculations of collision terms  
and the
new particles could contribute to $g_{\star}$ significantly. So our
results rely on the assumption that the new particle masses and $\Lambda$
are well above the temperatures relevant to our calculations. More
specifically, for the bounds presented in Fig.~\ref{fig:NMMNeff},
we are only concerned with $\nu_R$  decoupling at temperatures around or below
the electroweak scale. Therefore,  it is reasonable to assume that the new
physics scales are well above the decoupling temperature so that our results are independent
of the UV theories of NMM. 

Let us consider NMM generated at the one-loop level, for which the
magnitude of $\mu_{\nu}$ is roughly given by~\cite{Xu:2019dxe}
\begin{equation}
\mu_{\nu}\sim\frac{m_{X}}{16\pi^{2}\Lambda^{2}}\thinspace,\label{eq:NMM-th}
\end{equation}
where $16\pi^{2}$ is the one-loop suppression factor, $m_{X}$ denotes
the chirality-flipping fermion mass, and $\Lambda$ corresponds to
the heaviest particle mass in the loop. Taking the SM prediction $\mu_{\nu}=3eG_{F}m_{\nu}/(8\sqrt{2}\pi^{2})$
as an example, $\Lambda$ corresponds to the electroweak scale $G_{F}^{-1/2}$
and $m_{X}$ corresponds to $m_{\nu}$ because the chirality has to flip only  via the neutrino mass term\,---\,chirality-flipping via a charged-lepton mass is not feasible because $\nu_R$ does not participate in the SM gauge interactions.
In left-right symmetric models,
due to the presence of gauge interactions with $\nu_{R}$, chirality-flipping can be achieved by the charged fermion running in the loop~\cite{Shrock:1982sc}.
In this case, we would have $m_{X}=m_{\ell}$ with $\ell=e$, $\mu$,
or $\tau$, leading to a great enhancement of the NMM. Taking $\mu_{\nu}\sim10^{-12}\mu_{B}$
in Eq.~\eqref{eq:NMM-th}, we have $\Lambda\sim4.6\ \text{TeV}\cdot\sqrt{m_{X}/\text{GeV}}$,
which implies that for $m_{X}=m_{\tau}$ the new physics scale would
be well above TeV. For new heavy fermions playing the role of chirality
flipping, $\Lambda$ could be even higher.

\section{Conclusion}\label{sec:conclusion}
In this paper, we investigate cosmological constraints on Dirac NMM from current and future precision measurements of the effective number of neutrinos, $N_{\rm eff}$. As has been realized in previous studies, a straightforward calculation of the $\nu_R$ production rate using tree-level scattering amplitudes is IR divergent. Therefore, in this work we  compute the rate in a thermal QFT approach which, by taking  into account the effects of modified photon dispersion relations in the $\nu_R$ self-energy,  is free from the IR divergence.   
Furthermore, this approach automatically includes the contributions of the $s$- and $t$-channel scattering processes and plasmon decay.  


Using the refined $\nu_R$ production rate,
 we obtain accurate relations between $\mu_\nu$ and $N_{\rm eff}$, 
in which  the precision measurements of the latter can be recast to  constrain the former. 
The main results we have obtained  are presented in Fig.~\ref{fig:NMMNeff} and summarized in Tab.~\ref{tab:NMMbound}. For three $\nu_R$ being thermalized via flavor-universal NMM, the combination of current CMB and BBN measurements of $N_{\rm eff}$ puts a strong bound,  $\mu_{\nu}<2.7\times 10^{-12}\mu_B$ at $2\sigma$ C.L. This is better than the latest laboratory bounds from XENONnT and LZ. Future measurements from SO/SPT-3G and CMB-S4 will even be able to exclude scenarios with three or two $\nu_R$ being thermalized. 
Our results are applicable to a broad class of NMM models in which the new physic scale is much higher than the electroweak scale.

\appendix

\section{Weldon's formula with chiral fermions.\label{sec:weldon} }

In this appendix, we briefly review Weldon's formula~\cite{Weldon:1983jn}
which relates the gain/loss rate of a particle in the thermal bath
to its self-energy computed in finite-temperature field theories,
as we have applied in Eq.~\eqref{eq:totrate}. Although the formula
has been elaborated in great detail in the original paper, when applying
to chiral fermions, there could be a subtle issue regarding factors
of two. Hence we would like to take this appendix to clear potential
issues. 

\begin{figure}
	\centering
	
	\includegraphics[width=0.35\textwidth]{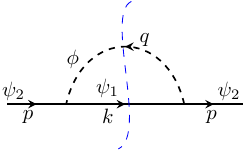}
	
	\caption{\label{fig:toy}The self-energy diagram of $\psi_{2}$ in the toy
		model.  }
	
\end{figure}

Let us first consider a non-chiral toy model:
\begin{equation}
{\cal L}\supset y\overline{\psi_{1}}\psi_{2}\phi+{\rm h.c.}\thinspace,\label{eq:A}
\end{equation}
where $\psi_{1,2}$ are two Dirac fermions and $\phi$ is a real scalar.
For simplicity, we   ignore the masses of $\psi_{1}$ and $\phi$,
and concentrate on the production and decay of $\psi_{2}$ in the
thermal bath. In this context, Weldon's formula (see Eq.~(2.31) in
Ref.~\cite{Weldon:1983jn}) reads
\begin{equation}
{\rm Im}\overline{u_{2}}\Sigma u_{2}=-E_{p}\Gamma_{{\rm tot}}\thinspace,\label{eq:A-1}
\end{equation}
where $u_{2}$ ($\overline{u}_{2}$) denotes the initial (final) state
of the $\psi_{2}$ particle, $\Sigma$ denotes its self-energy shown
in Fig.~\ref{fig:toy}, $E_{p}$ and $p$ are the particle energy
and momentum, and $\Gamma_{{\rm tot}}\equiv\Gamma_{{\rm gain}}+\Gamma_{{\rm loss}}$
with $\Gamma_{{\rm gain}}$ and $\Gamma_{{\rm loss}}$ defined similar
to those in Eq.~\eqref{eq:f-boltz}.

Below we explicitly verify Eq.~\eqref{eq:A-1} by computing both sides
independently.

The self-energy on the left-hand side reads (see Eq.~(2.21) in Ref.~\cite{Weldon:1983jn}):
\begin{equation}
\Sigma(p)=-y^{2}T\sum_{b=-\infty}^{\infty}\int\frac{d^{3}k}{(2\pi)^{3}}S_{\psi_{1}}(k)D_{\phi}(q)\thinspace,\label{eq:A-2}
\end{equation}
where the momenta $p$, $k$, $q$ have been specified in Fig.~\ref{fig:toy},
$S_{\psi_{1}}$ and $D_{\phi}$ denote the propagators of $\psi_{1}$
and $\phi$ at a finite temperature, $T$. Following Ref.~\cite{Weldon:1983jn},
here we use the imaginary-time formalism, in which the energy take
discrete values $i2\pi Tb$ where $b$ is an integer for a boson or
a half-integer for a fermion. For this reason, the energy integral
$\int dk^{0}/(2\pi)$ has been discretized to the summation of $b$.
After computing the summation, one can get 
\begin{equation}
{\rm Im}\overline{u_{2}}\Sigma u_{2}=-\frac{1}{2}y^{2}\int\frac{d^{3}k}{(2\pi)^{3}}\frac{2\pi}{2E_{k}2E_{q}}\left[\delta(E_{p}-E_{k}-E_{q})\overline{u_{2}}\slashed{k}u_{2}(1-f_{\psi_{1}}+f_{\phi})+\cdots\right],\label{eq:A-3}
\end{equation}
where ``$\cdots$'' represents other terms proportional to $\delta(E_{p}+E_{k}-E_{q})$,
$\delta(E_{p}-E_{k}+E_{q})$, etc. We refer to Eq.~(2.22) in Ref.~\cite{Weldon:1983jn}
for their explicit forms. 

Now let us turn to the right-hand side of Eq.~\eqref{eq:A-1}. Recall
that the collision term in the Boltzmann equation for $f_{\psi_{2}}$
is given by
\begin{align}
C[f_{\psi_{2}}] & =\frac{1}{2E_{p}}\left[\sum_{s_{1}}\int d\Pi_{\psi_{1}}d\Pi_{\phi}(2\pi)^{4}\delta{}^{4}f_{\psi_{1}}f_{\phi}(1-f_{\psi_{2}})|{\cal M}(\psi_{1}\phi\to\psi_{2})|^{2}\right.\nonumber \\
& -\left.\int d\Pi_{\psi_{1}}d\Pi_{\phi}(2\pi)^{4}\delta{}^{4}f_{\psi_{2}}(1+f_{\phi})(1-f_{\psi_{1}})\sum_{s_{1}}|{\cal M}(\psi_{2}\to\psi_{1}\phi)|^{2}\right],\label{eq:A-4}
\end{align}
where $s_{1}$ denotes the spin of $\psi_{1}$ and $d\Pi_{x}\equiv d^{3}p_{x}/\left[(2\pi)^{3}2E_{x}\right]$
does not include internal degrees of freedom. Instead, we have $\sum_{s_{1}}$
to sum over the spin of $\psi_{1}$ explicitly. The squared amplitudes
read 
\begin{equation}
|{\cal M}(\psi_{1}\phi\to\psi_{2})|^{2}={\cal M}(\psi_{2}\to\psi_{1}\phi)|^{2}=y^{2}|\overline{u}_{1}u_{2}|^{2}=y^{2}|\overline{u}_{2}u_{1}|^{2}\thinspace.\label{eq:A-6}
\end{equation}
After the spin summation, we have 
\begin{equation}
\sum_{s_{1}}|{\cal M}|^{2}=\sum_{s_{1}}y^{2}{\rm tr}\left[u_{1}^{s_{1}}\overline{u}_{1}^{s_{1}}u_{2}\overline{u}_{2}\right]=y^{2}\overline{u}_{2}\slashed{k}u_{2}\thinspace,\label{eq:A-7}
\end{equation}
which is identical to the part between $\delta(E_{p}-E_{k}-E_{q})$
and $(1-f_{\psi_{1}}+f_{\phi})$ in Eq.~\eqref{eq:A-3}. 

By definition, we have $C[f_{\psi_{2}}]=(1-f_{\psi_{2}})\Gamma_{{\rm gain}}-f_{\psi_{2}}\Gamma_{{\rm loss}}$.
Then Eq.~\eqref{eq:A-4} implies
\begin{align}
\Gamma_{{\rm tot}} & =\frac{y^{2}}{2E_{p}}\int d\Pi_{\psi_{1}}d\Pi_{\phi}(2\pi)^{4}\delta{}^{4}\left[f_{\psi_{1}}f_{\phi}+(1+f_{\phi})(1-f_{\psi_{1}})\right]\overline{u}_{2}\slashed{k}u_{2}\nonumber \\
& =\frac{y^{2}}{2E_{p}}\int\frac{d^{3}k}{(2\pi)^{3}}\frac{2\pi}{2E_{k}2E_{q}}\delta(E_{p}-E_{k}-E_{q})\left[1-f_{\psi_{1}}+f_{\phi}\right]\overline{u}_{2}\slashed{k}u_{2},\label{eq:A-5}
\end{align}
where in the second step we have integrated out $d\Pi_{\phi}$ with
a part of the delta function. 

By comparing Eq.~\eqref{eq:A-5} with Eq.~\eqref{eq:A-3}, we see
that Weldon's formula is explicitly verified, except for the ``$\cdots$''
part in Eq.~\eqref{eq:A-3} accounting for contributions of other
processes. 

Note that so far we have not summed the spin of $\psi_{2}$, nor have
we included its internal degrees of freedom in the calculation. Hence
both sides of Eq.~\eqref{eq:A-1} should be interpreted as quantities
for only one of its degrees of freedom (if there are many). For the
non-chiral model considered above, since the results for the two possible
spin polarizations of a $\psi_{2}$ particle are equal, one can average
over the spin polarizations:
\begin{equation}
\Gamma_{{\rm tot}}=-\frac{{\rm Im}\overline{u_{2}}\Sigma u_{2}}{E_{p}}=-\frac{1}{2}\sum_{s_{2}}\frac{{\rm Im}\overline{u_{2}^{s_{2}}}\Sigma u_{2}^{s_{2}}}{E_{p}}=-\frac{{\rm tr}\left[(\slashed{p}+m_{2})\Sigma\right]}{2E_{p}}\thinspace.\label{eq:A-8}
\end{equation}

Now let us revisit the calculation for the following chiral toy model:
\begin{equation}
{\cal L}\supset y\overline{\psi_{1R}}\psi_{2L}\phi+{\rm h.c.}\thinspace,\label{eq:A-9}
\end{equation}
i.e.,~we assume that only the left-handed part of $\psi_{2}$ participates in 
the yukawa interaction. For this model, one can still explicitly verify
Weldon's formula with a few modifications as follows. First, the two
vertices in Fig.~\ref{fig:toy} are accompanied with the chiral projectors
$P_{L}$ and $P_{R}$. So we have $S_{\psi_{1}}(k)\to P_{R}S_{\psi_{1}}(k)P_{L}$
in Eq.~\eqref{eq:A-2} and $\overline{u_{2}}\slashed{k}u_{2}\to\overline{u_{2}}P_{R}\slashed{k}P_{L}u_{2}$
in Eq.~\eqref{eq:A-3}. The squared amplitude is also modified:
\begin{equation}
|{\cal M}|^{2}\to y^{2}|\overline{u}_{1}P_{L}u_{2}|^{2}=y^{2}|\overline{u}_{2}P_{R}u_{1}|^{2}\thinspace.\label{eq:A-10}
\end{equation}
Here the projectors automatically guarantee that only left-handed
$\psi_{2}$ and right-handed $\psi_{1}$ are involved in the amplitude.
So one can still sum over the spin because the contribution of the
wrong spin polarization automatically vanish due to $P_{L/R}$. 

With the above details being noted, we can see that Eq.~\eqref{eq:A-1}
for chiral fermions still holds. However, due to the absence of right-handed
$\psi_{2}$, Eq.~\eqref{eq:A-8} should be modified as 
\begin{equation}
\Gamma_{{\rm tot}}=-\frac{{\rm Im}\overline{u_{2}}P_{R}\Sigma P_{L}u_{2}}{E_{p}}=-\sum_{s_{2}}\frac{{\rm Im}\overline{u_{2}^{s_{2}}}P_{R}\Sigma P_{L}u_{2}^{s_{2}}}{E_{p}}=-\frac{{\rm tr}\left[(\slashed{p}+m_{2})P_{R}\Sigma P_{L}\right]}{E_{p}}\thinspace,\label{eq:A-11}
\end{equation}
where in principle the projectors $P_{L/R}$ should be included in
$\Sigma$ but here we prefer to write them out explicitly. For our
application to neutrinos in Sec.~\ref{sec:Tcomputation}, the projectors have been included
implicitly in Eq.~\eqref{eq:totrate}. 

\section{A consistent check for the  retarded amplitude}\label{sec:check}
Ref.~\cite{Elmfors:1997tt} mentioned that the imaginary part of the neutrino self-energy in its Eq.~(4.2)  was time-ordered. The explicit form is given as follows:
\begin{align}\label{eq:ImSigma}
	\text{Im}\Sigma(p)&=-\frac{\mu^2\epsilon(p_0)}{\sin2\phi_p}\int \frac{d^4 k }{(2\pi)^4}\epsilon(p_0+k_0)\epsilon(k_0)\frac{1}{2}\sin2\phi_{p+k}\frac{1}{2}\sinh2\theta_k
	\nonumber\\[0.2cm]
	&\times k_\alpha \sigma^{\alpha \mu}(\slashed{p}+\slashed{k})\frac{1}{2}(1-\gamma_5)k_\beta \sigma^{\beta\nu}(2\pi)^2\delta[(p+k)^2]\mathcal{A}_{\mu\nu}(k)\,,
\end{align}
where $\epsilon(x)$ is the sign function of $x$ and 
\begin{align}
	\frac{1}{2}\sin2\phi_{k}=\frac{e^{|k_0|/(2T)}}{e^{|k_0|/T}+1}\, ,\quad \frac{1}{2}\sinh2\theta_k=\frac{e^{|k_0|/(2T)}}{e^{|k_0|/T}-1}\, .
\end{align}
The photon spectral function is defined by $\mathcal{A}_{\mu\nu}=-P_{\mu\nu}\mathcal{A}_T-Q_{\mu\nu}\mathcal{A}_L$, with $\mathcal{A}_{T,L}$ given by   Eq.~(3.9) in Ref.~\cite{Elmfors:1997tt} which, up to a minus sign  convention, is   identical    to  Eq.~\eqref{eq:spectraldensity} for non-vanishing $\text{Im}\Pi^{T,L}_R$. Substituting Eq.~\eqref{eq:ImSigma} into the total decay rate $\Gamma_{\rm tot}$, the authors obtained 
\begin{align}\label{eq:GammaR}
	\Gamma_{\rm tot}&=\frac{\mu^2}{2\pi}\int_0^\infty |\vec k| d|\vec{k}| \int_{-\infty}^{\infty}dk_0 \theta\left(-k^2[(2|\vec p|+k_0)^2-|\vec k|^2]\right)
	\nonumber \\[0.2cm]
	&\times\left[\epsilon(k_0)f_{\nu_L}(|k_0+p_0|)+\epsilon(p_0+k_0)f_\gamma(|k_0|)+\epsilon(p_0+k_0)\theta(-k_0)-\epsilon(k_0)\theta(-p_0-k_0)\right]
		\nonumber \\[0.2cm]
	&\times\frac{k^4}{4|\vec k|^2 |\vec{p}|^2}\left[((2|\vec{p}|+k_0)^2-|\vec{k}|^2)\mathcal{A}_T-(2|\vec{p}|+k_0)^2\mathcal{A}_L\right]\epsilon(k_0)\, ,
\end{align}
 as given by Eq.~(4.4) in Ref.~\cite{Elmfors:1997tt}. To make a comparison with our results shown in Eqs.~\eqref{eq:totrate} and~\eqref{trace-sum}, we need the following identities:
 \begin{align}
 	\epsilon(k_0)f_{\nu_L}(|k_0+p_0|)&=\epsilon(k_0)[\epsilon(k_0+p_0) f_{\nu_L}(p_0+k_0)+\theta(-p_0-k_0)]
 	\nonumber\\[0.2cm]
 	&=\epsilon(k_0)\epsilon(p_0+k_0)f_{\nu_L}(p_0+k_0)+\epsilon(k_0)\theta(-p_0-k_0) \, ,
 \\[0.2cm]
 \epsilon(p_0+k_0)f_{\gamma}(|k_0|)&=\epsilon(k_0+p_0)[\epsilon(k_0)f_\gamma(k_0)-\theta(-k_0)]
 	\nonumber\\[0.2cm]
 &=\epsilon(p_0+k_0)\epsilon(k_0)f_\gamma(k_0)-\epsilon(p_0+k_0)\theta(-k_0)\, ,
 \end{align}
where $f(|x|)=\theta(x)f(x)+\theta(-x)f(-x)$, $\theta(x)-\theta(-x)=\epsilon(x)$, $f_{\nu_L}(x)+f_{\nu_L}(-x)=1$ and $f_{\gamma}(x)+f_{\gamma}(-x)=-1$ have been used. 
  Then Eq.~\eqref{eq:GammaR} is reduced to 
  \begin{align}\label{eq:Gammatot2}
  	\Gamma_{\rm tot}&=\frac{\mu^2}{2\pi}\int_0^\infty |\vec k| d|\vec{k}| \int_{-\infty}^{\infty}dk_0 \theta\left(-k^2[(2|\vec p|+k_0)^2-|\vec k|^2]\right)
  \nonumber \\[0.2cm]
  &\times \epsilon(p_0+k_0)[f_{\nu_L}(p_0+k_0)+f_\gamma(k_0)]
  \nonumber \\[0.2cm]
  &\times\frac{k^4}{4|\vec k|^2 |\vec{p}|^2}\left[((2|\vec{p}|+k_0)^2-|\vec{k}|^2)\mathcal{A}_T-(2|\vec{p}|+k_0)^2\mathcal{A}_L\right]\, .
  \end{align}
  The remaining $\theta$-function and sign function in the above result is given by
  \begin{align}\label{eq:thetapro}
  &	\theta\left(-k^2[(2|\vec p|+k_0)^2-|\vec k|^2]\right) \epsilon(p_0+k_0)
  	\nonumber \\[0.2cm]
  	&=\left(\theta(k^2)\theta[|\vec k|^2-(2p_0+k_0)^2]+\theta(-k^2)\theta[(2p_0+k_0)^2-|\vec{k}|^2]\right)[\theta(p_0+k_0)-\theta(-p_0-k_0)]
  	\nonumber \\[0.2cm]
  	&=\left(\theta(-k_0-|\vec{k}|)\theta[|\vec{k}|^2-(2p_0+k_0)^2]+\theta(|\vec{k}|^2-k_0^2)\theta(2p_0+k_0-|\vec{k}|)\right)
  	\nonumber \\[0.2cm]
  	&\times[\theta(p_0+k_0)-\theta(-p_0-k_0)]\, ,
  \end{align}
where we have used the fact that right-handed neutrinos are relativistic $|\vec{p}|=p_0$.
  Note that $\theta(-k_0-|\vec{k}|)\theta[|\vec{k}|^2-(2p_0+k_0)^2]$ indicates that $-|\vec{k}|<2p_0+k_0<|\vec{k}|$ and $k_0+|\vec{k}|<0$. Then $k_0-|\vec{k}|<2p_0+2k_0<k_0+|\vec{k}|$ indicates that  $p_0+k_0<0$. Similarly, $\theta(|\vec{k}|^2-k_0^2)\theta(2p_0+k_0-|\vec{k}|)$ indicates that $2p_0+k_0>|\vec{k}|$ and $-|\vec{k}|<k_0<|\vec{k}|$, so  this $\theta$ product indicates that  $p_0+k_0>0$. Consequently, Eq.~\eqref{eq:thetapro}   reduces to $-\Theta_1+\Theta_2$, where $\Theta_{1,2}$ are given by Eqs.~\eqref{eq:Theta1}-\eqref{eq:Theta2}. Putting this result back to Eq.~\eqref{eq:Gammatot2},  we can see that it coincides with Eq.~\eqref{eq:totrate} in which the trace is given by Eq.~\eqref{trace-sum}. Therefore, the amplitude used in Ref.~\cite{Elmfors:1997tt} is in fact  retarded  rather than  time-ordered as claimed by the authors.
  
\section*{Acknowledgements}
This work is supported in part by the National Natural Science Foundation of China under grant No. 12141501. 

\bibliographystyle{JHEP}
\bibliography{AllRef}

\end{document}